\documentclass[11pt]{article}
\usepackage{color}
\usepackage[normalem]{ulem}
\usepackage{tikz}
\usepackage{todonotes}
\usepackage{epsfig,enumerate,float,amsmath,amsfonts,amsbsy,amssymb,amsthm,mathrsfs,lineno,ifpdf,relsize,float,etoolbox}
\usepackage{xcolor}
\usepackage[colorlinks=true, allcolors=blue]{hyperref}
\usepackage[ruled,vlined,linesnumbered,procnumbered]{algorithm2e}
\usepackage[mathscr]{euscript}
\usepackage{graphicx}
\usepackage{setspace}
\usepackage{makecell}
\usepackage[utf8]{inputenc}
\usepackage[english]{babel}
\usepackage[numbers]{natbib}
\usepackage[usenames,dvipsnames]{pstricks}
\usepackage{pst-grad}
\usepackage{float}
\usepackage{tikz}
\usetikzlibrary{positioning,fit,backgrounds,arrows.meta}
\theoremstyle{plain}
\newtheorem{theorem}{Theorem}
\newtheorem{lemma}{Lemma}

\newtheorem{coro}{Corollary}
\newtheorem{obs}{Observation}
\newtheorem{propo}{Proposition}

\newtheorem{claim}{Claim}

\newtheorem{defi}{Definition}

\makeatletter
\AtBeginEnvironment{procedure}{\let\c@algocf\c@procedure}
\makeatother

\let\oldenumerate\enumerate
\renewcommand{\enumerate}{
	\oldenumerate
	\setlength{\itemsep}{1.5pt}
	\setlength{\parskip}{0pt}
	\setlength{\parsep}{0pt}
}
\newcommand{\gds}{\gamma_{ds}}

\title{Complexity and algorithms for dual-server domination in graphs}
\author{$^1$Sumanta Banerjee, $^2$Dinabandhu Pradhan, $^2$Vaishali Sharma, $^3$Shaily Verma\\ \\
	$^{1}$Department of Mathematics\\ Silda Chandrasekhar College, India\\
	$^{2}$Department of Mathematics \& Computing\\ Indian Institute of Technology (ISM), Dhanbad\\
	$^{3}$Department of Mathematics\\ Indian Institute of Technology, Jodhpur, India\\
	\small \tt Email: sumanta.banerjee5@gmail.com; dina@iitism.ac.in;\\ vaishali.sharma7791@gmail.com; shailyverma@iitj.ac.in}

\begin{document}

	\maketitle
	\begin{abstract}
		A set $D\subseteq V(G)$ is a \emph{dual-server dominating set} if $D$ can be partitioned into two disjoint sets $R$ and $B$ such that every vertex $v\in V(G)\setminus D$ has a neighbor in both $R$ and $B$. Intuitively, one can think of $R$ and $B$ as two different service providers, where every vertex outside $D$ must have access to both services. The minimum cardinality of a dual-server dominating set is the \emph{dual-server domination number} of $G$ and is denoted by $\gamma_{{ds}}(G)$. Given a graph $G$ and an integer $k>0$, \textsc{DualServer} is the problem of deciding whether $G$ has a dual-server dominating set of cardinality at most $k$.

		\medskip
		In this paper, we study the computational complexity and algorithmic aspects of the minimum dual-server domination problem. Chellali et al. [Discrete Appl. Math. 387 (2026) 251-259] proved that \textsc{DualServer} is NP-complete on bipartite graphs. We establish NP-completeness for \textsc{DualServer} on chordal bipartite graphs, planar graphs, undirected path graphs, and split graphs. We further prove that, for every fixed $0<\varepsilon<1$, unless P=NP, the problem of computing dual-server domination number does not admit any polynomial-time approximation algorithm with approximation ratio \(n^{1-\varepsilon}\), even on bipartite graphs, where $n$ denotes the number of vertices. On the algorithmic side, we show that \(\gamma_{ds}(G)=\gamma_2(G)\) for threshold graphs, chain graphs, block graphs, and proper interval graphs. For threshold and chain graphs, we derive exact formulas for the dual-server domination number, which in turn yields linear time algorithms. We also present linear time algorithms for $P_4$-sparse graphs and two-terminal series-parallel graphs.
	\end{abstract}
	

	\section{Introduction}
	\label{sec:introduction}
	
	All the graphs considered in this paper are finite, connected, simple, and undirected. We follow the notations and terminologies used in \cite{textBook}. We denote the set $\{1, 2, \ldots, t\}$ by $[t]$. Domination is a fundamental problem in graph theory with numerous applications. A set $D\subseteq V(G)$ is a \emph{dominating set} of a graph $G$ if every vertex $v\in V(G)\setminus D$ has a neighbor in $D$. Domination provides a natural model for selecting service providers in a network. One can think vertices in $D$ represent service providers, and every vertex outside $D$ has access to a service provider. When two distinct services are
	required, the selected vertices must also be assigned service types so that every vertex outside the selected set has access to both. Motivated by this requirement, Chellali et al.~\cite{Chellali2026} recently introduced a new variation of domination, called \emph{dual-server domination}. A set $D\subseteq V(G)$ is a \emph{dual-server dominating set} if $D$ can be partitioned into two disjoint sets $R$ and $B$ such that every vertex $v\in V(G)\setminus D$ has a neighbor in both $R$ and $B$. Intuitively, one can think of $R$ and $B$ as two different service providers, where every vertex outside $D$ must have access to both services. The minimum cardinality of a dual-server dominating set is the \emph{dual-server domination number} of $G$ and is denoted by $\gamma_{{ds}}(G)$. The associated decision problem is defined as follows.

	\begin{center}
		\fbox{
			\begin{minipage}{0.9\linewidth}
				\underline{\textsc{DualServer}}
				
				\vspace{0.5em}
				
				\textbf{Input:} A graph $G = (V, E)$ and an integer $k>0$.
				
				\textbf{Question:} Does $G$ have a dual-server dominating set of cardinality at most $k$?
			\end{minipage}
		}
	\end{center}

	A set $S\subseteq V(G)$ is a \emph{$2$-dominating set} if each vertex $v\in V(G)\setminus S$ has at least two neighbors in $S$. The minimum cardinality of a $2$-dominating set is the \emph{$2$-domination number} of $G$ and is denoted by $\gamma_{2}(G)$. The following observation is due to the fact that every dual-server dominating set is a $2$-dominating set.
	
	\begin{obs}[\cite{Chellali2026}]\label{prop:twodom}
		$\gamma_2(G)\le\gamma_{ds}(G)$.
	\end{obs}  
	
\begin{figure}[h]
		\begin{center}
			\includegraphics[width=0.5\linewidth]{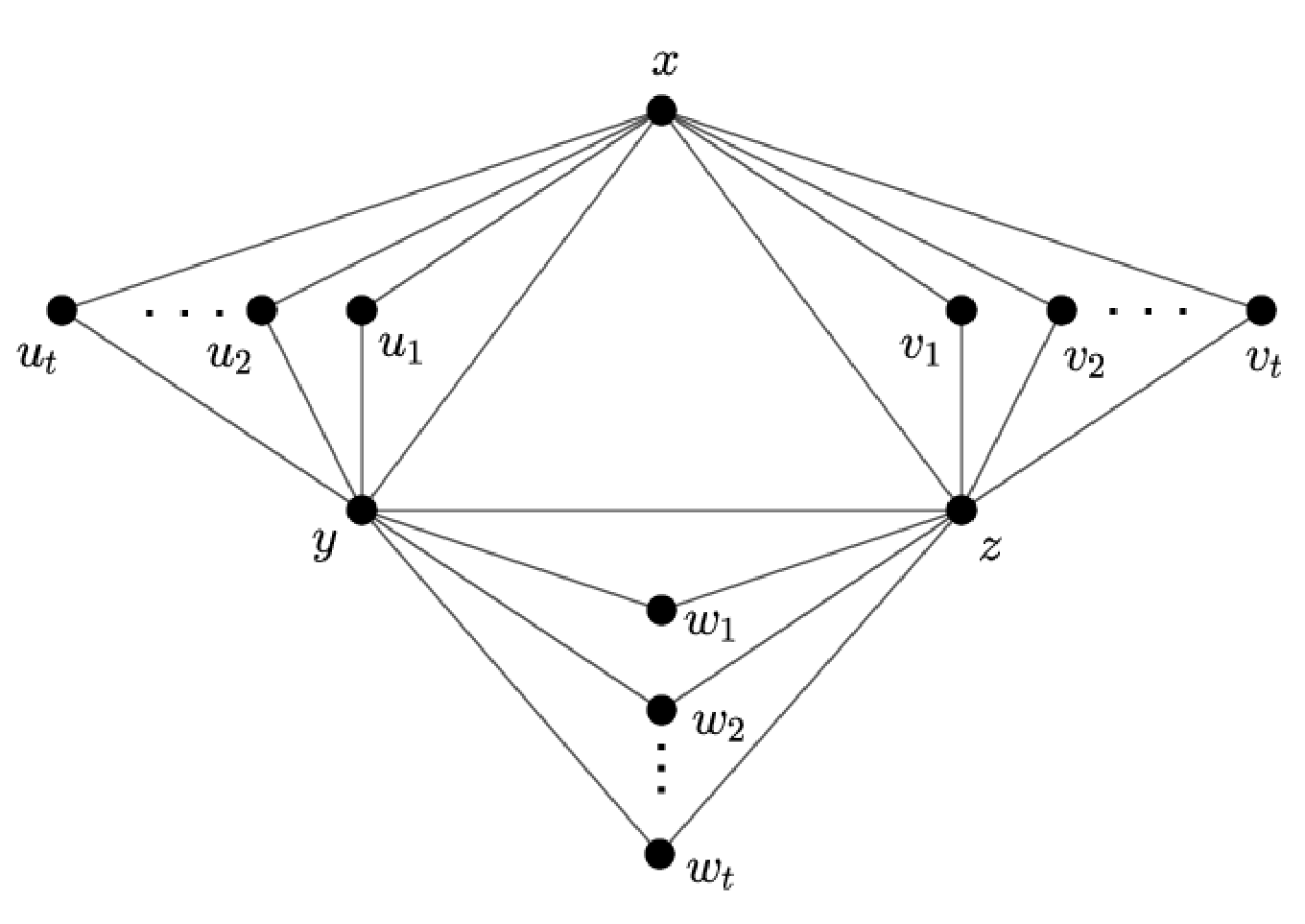}
			\caption{The graph $\mathcal{H}_t$.}
			\label{large_diff_2-ds_figure}
		\end{center}	
	\end{figure}

	However, a $2$-dominating set may not be a dual-server dominating set. There are graphs for which the equality holds. For example, $\gamma_{2}(K_n)=\gamma_{ds}(K_n)$. In this paper, we show that $\gamma_{2}(G)=\gamma_{{ds}}(G)$ for every block graph $G$. In fact, there are other graph classes for which the equality holds. Despite the equality in some classes, it is interesting to see that $\gamma_{{ds}}(G)-\gamma_{2}(G)$ can be arbitrarily large. For each integer $t\geq 2$, construct a graph $\mathcal{H}_t$ as follows. Let $\{x,y,z\}$ be a clique, and let $U,V,W$ be pairwise disjoint independent sets of cardinality $t$, with no edges between these sets. Every vertex of $U$, $V$, and $W$ has neighborhood $\{x,y\}$, $\{x,z\}$, and $\{y,z\}$, respectively. See Figure~\ref{large_diff_2-ds_figure}. Then, $\gamma_2(\mathcal{H}_t)=3$ and $\gamma_{ds}(\mathcal{H}_t)=t+3$. Therefore, the difference between $\gamma_{ds}(\mathcal{H}_t)$ and $\gamma_2(\mathcal{H}_t)$ is unbounded.
	
	\subsection{Related work}

	Chellali et al.~\cite{Chellali2026} studied
	basic properties of dual-server domination and compared it with other domination parameters. They proved that
	$\gamma_{ds}(T)=\gamma_2(T)$ for every tree $T$, which yields a linear time algorithm for computing the dual-server domination number of a tree. They also characterized connected graphs attaining large values of $\gamma_{{ds}}(G)$, including graphs for which $\gamma_{{ds}}(G)=|V(G)|-1$ and $\gamma_{{ds}}(G)=|V(G)|-2$.
	On the complexity side, they proved that \textsc{DualServer} is NP-complete on bipartite graphs and chordal graphs. They concluded by asking for graph families on which the dual-server domination number can be computed in polynomial time.
	Rao et al.~\cite{rao2026} determined the dual-server domination number of the splitting graphs of paths, cycles, complete graphs, and complete bipartite graphs. Moreover, they studied the dual-server domination number under various graph operations including the join, corona, and Cartesian product of graphs.

	\subsection{Our contributions}
	
	We study dual-server domination in different graph classes.
	Our main results are as follows.
	
	\begin{itemize}
		
		\item \textbf{Complexity on graph classes.}
		We prove that \textsc{DualServer} is NP-complete on chordal bipartite graphs, strengthening the previously established hardness on bipartite graphs~\cite{Chellali2026}. Furthermore, we establish NP-completeness for \textsc{DualServer} on planar graphs, undirected path graphs, and split graphs.
		
		\item \textbf{Hardness of approximation.}
		For every fixed $0<\varepsilon<1$, we prove that, unless
		P=NP, the dual-server domination number cannot be approximated in polynomial time within a factor of $n^{1-\varepsilon}$, even on bipartite graphs, where $n$ is the number of vertices of the input graph. Several classical domination variants, including domination, total domination, Roman domination, and $2$-domination, admit logarithmic approximation algorithms and have logarithmic inapproximability bounds~\cite{BonomoEtAl2018,CMV2013,PoureidiFathali2023}.
		In particular, minimum $2$-domination admits a polynomial-time $(\ln(\Delta+2)+1)$-approximation algorithm, where $\Delta$ is the maximum degree~\cite{BonomoEtAl2018}.
		Thus, the requirement to assign two server types creates a
		substantial difference in approximability between
		dual-server domination and ordinary $2$-domination. Similar inapproximability bounds have been established for very few variations of domination. In particular, perfect Italian domination~\cite{PradhanBanerjeeLiu2022} and unique response Roman domination~\cite{BanerjeeChaudharyPradhan2023} admits similar $n^{1-\varepsilon}$ inapproximability bound.

		\item \textbf{Relationship with $2$-domination.}
		We give constructive proofs of
		$\gamma_{ds}(G)=\gamma_2(G)$ for threshold graphs,
		chain graphs, block graphs, and proper interval graphs.
		These equalities show that assigning two server types
		introduces no additional cost in the optimum on these
		classes.

		\item \textbf{Linear time algorithms.}
		We derive exact formulas for the dual-server domination
		number of threshold graphs and chain graphs, yielding
		linear time algorithms for computing the dual-server domination number. We also give linear time algorithms for computing the dual-server domination number of $P_4$-sparse graphs and two-terminal series-parallel graphs, using their recursive decompositions. Note that $P_4$-sparse graphs contain the class of threshold graphs.
		The equality with $2$-domination need not hold on
		$P_4$-sparse and series-parallel graphs. The graph
		$H=2K_2+2K_2$ is $P_4$-free, and hence $P_4$-sparse,
		but satisfies $\gamma_2(H)=3$ and $\gamma_{ds}(H)=4$. The graph $G=C_6$ is series parallel and satisfies $\gamma_2(G)=3$ and $\gamma_{ds}(G)=4$.
	\end{itemize}

	We conclude this section with the following auxiliary result, which will be used in establishing our subsequent results.
	
	\begin{obs}\label{pendant_observation_ds}
		If $v\in V(G)$ is a vertex with $d(v)\le1$, then $v$ must be in each dual-server dominating set and each $2$-dominating set of $G$.
	\end{obs}

	\section{NP-completeness}\label{np}
	
	In this section, we prove that \textsc{DualServer} is NP-complete for chordal bipartite graphs, planar graphs, undirected path graphs, and split graphs. A graph $G = (V, E)$ is called an \emph{intersection graph} for a finite family $\mathcal{F}$ of a non-empty set if there is a one-to-one correspondence between $\mathcal{F}$ and $V$ such that two sets in $\mathcal{F}$ have nonempty intersection if and only if their corresponding vertices are adjacent in $V$. A graph $G$ is said to be an \emph{undirected path graph} if $G$ is the intersection graph of a family of paths of a tree. A graph $G$ is \emph{chordal bipartite} if $G$ is bipartite and each cycle of $G$ of length greater than $4$ has a chord. A graph $G$ is said to be \emph{planar} if it can be drawn in the plane such that no two edges intersect except possibly at their common endpoints.

	We reduce an instance of \textsc{Dominating set} to an instance of \textsc{DualServer} in polynomial time. Let $G$ be a with $V(G)=\{v_1, v_2, \ldots, v_n\}$. We construct a graph $G'$ from the graph $G$ in polynomial time using the following steps. 
	
	\begin{enumerate}
		\item Initialize $G'$ as a copy of $G$. Let $V_1=\{v_1, v_2, \ldots, v_n\}$.
		\item Attach a pendant vertex $x_i$ to each vertex $v_i\in V_1$. Let $V_2=\{x_1, x_2, \ldots, x_n\}$.
	\end{enumerate}

	\begin{lemma}\label{SO_npc}
		$G$ has a dominating set of cardinality $k$ if and only if $G'$ has a dual-server dominating set of cardinality $k+n$.
	\end{lemma}
	\begin{proof}
		
		Let $G$ has a dominating set $S$ of cardinality $k$. Let $D= R\cup B$ be such that $R=V_2$ and $B=S$. It is easy to see that $|D|=k+n$. Now, we prove that $D$ is a dual-server dominating set. Since $S$ is a dominating set of $G$ and every vertex of $V_2$ is in $D$, every vertex of $V(G') \setminus D$ has a neighbor in $B$ and a neighbor in $R$. Therefore, every vertex $v\in V(G')\setminus D$ has a neighbor in $R$ and a neighbor in $B$. Thus, $D$ is a dual-server dominating set.
		
		\medskip
		
		Let $G'$ has a dual-server dominating set $D$ of cardinality $k+n$. By Observation~\ref{pendant_observation_ds}, each vertex of $V_2$ is in $D$. Let $S=D\cap V_1$. Since $D$ is a dual server dominating set and every vertex of $V_1$ has exactly one neighbor in $D\cap V_2$, each vertex of $V_1\setminus S$ has a neighbor in $S$. Therefore, $S$ is a dominating set of $G$.
	\end{proof}
	
	Note that $G'$ is a chordal bipartite graph when $G$ is a chordal bipartite graph, and $G'$ is a planar graph when $G$ is a planar graph. Let $G$ be an undirected path graph. Then, there is a tree $T$ and a set of paths $\{P_{v_i} : 1 \leq i \leq n\}$ of $T$ such that $v_iv_j \in E$ if and only if
	$P_{v_i} \cap P_{v_j} \neq \emptyset$. For each path $P_{v_i}$, consider an end point $v_i^*$ of $P_{v_i}$. Construct a tree $T'$ which results from
	$T$ by joining a new path $v_i^*, x_i$ of length
	$1$ at $v_i^*$ (i.e., $P_{v_i}$ is extended to $P_{v_i}^*$, where $P_{v_i}^* = P_{v_i} \cup \{v_i^*x_i\}$ for each $v_i$, $1 \leq i \leq n$. Then it is clear that $G'$ is the intersection graph of the set of paths $\bigcup_{v_i \in V}
	\left\{
	P_{v_i} \cup \{v_i^*x_i\}
	\right\}$. Hence $G'$ is an undirected path graph when $G$ is an undirected path graph.
	
	\textsc{Dominating Set} is known to be NP-complete for chordal bipartite graphs~\cite{Muller}, for planar graphs~\cite{Kikuno1980}, and for undirected path graphs~\cite{BoothJohnson1982}. Therefore, by Lemma~\ref{SO_npc}, we have the following theorem. 
	
	\begin{theorem}\label{Dserver:np_complete}
		\textsc{DualServer} is NP-complete in chordal bipartite graphs, in planar graphs, and in undirected path graphs.   
	\end{theorem}

	\subsection{Split graphs}
	
	A graph $G$ is a \emph{split graph} if $V(G)$ can be partitioned into two sets $C$ and $I$ where $C$ is a clique and $I$ is an independent set. It is known that \textsc{Vertex Cover} problem is NP-complete for general graphs~\cite{GJ}.

	\begin{theorem}\label{split:np_complete}
		\textsc{DualServer} is NP-complete in split graphs.
	\end{theorem}
	\begin{proof}
		We reduce an instance of \textsc{Vertex Cover} in general graphs to an instance of \textsc{DualServer} in split graphs in polynomial time. Let $G$ be a graph with $V(G)=\{v_1, v_2,\ldots, v_n\}$ and $E(G)=\{e_1, e_2, \ldots, e_m\}$. We construct a graph $G'$ as shown in Figure~\ref{np_split_figure} from the graph $G$ in polynomial time using the following steps. 
		
		\begin{enumerate}
			\item Let $V_1=V(G)$. Make $V_1$ a clique.
			
			\item Let $V_2=\{v_{e_i}~|~e_i\in E(G)\}$. For each $k\in[m]$ and $i,j\in[n]$, $v_{e_k}$ is adjacent to $v_i$ and $v_j$ if and only if $v_i$ and $v_j$ are the endpoints of $e_k$ in $G$.

			\item Attach a vertex $x_1$ to each vertex of $V_1$ and a vertex $x_2$ to each vertex of $V_1$ and $V_2$.
			
			\item Add an edge between $x_1$ and $x_2$.
			
			\item Add three vertices $u_1, u_2, u_3$ and make each of them adjacent to $x_1$ and $x_2$.
		\end{enumerate}

		\begin{figure}[h]
			\begin{center}
				\includegraphics[width=0.5\linewidth]{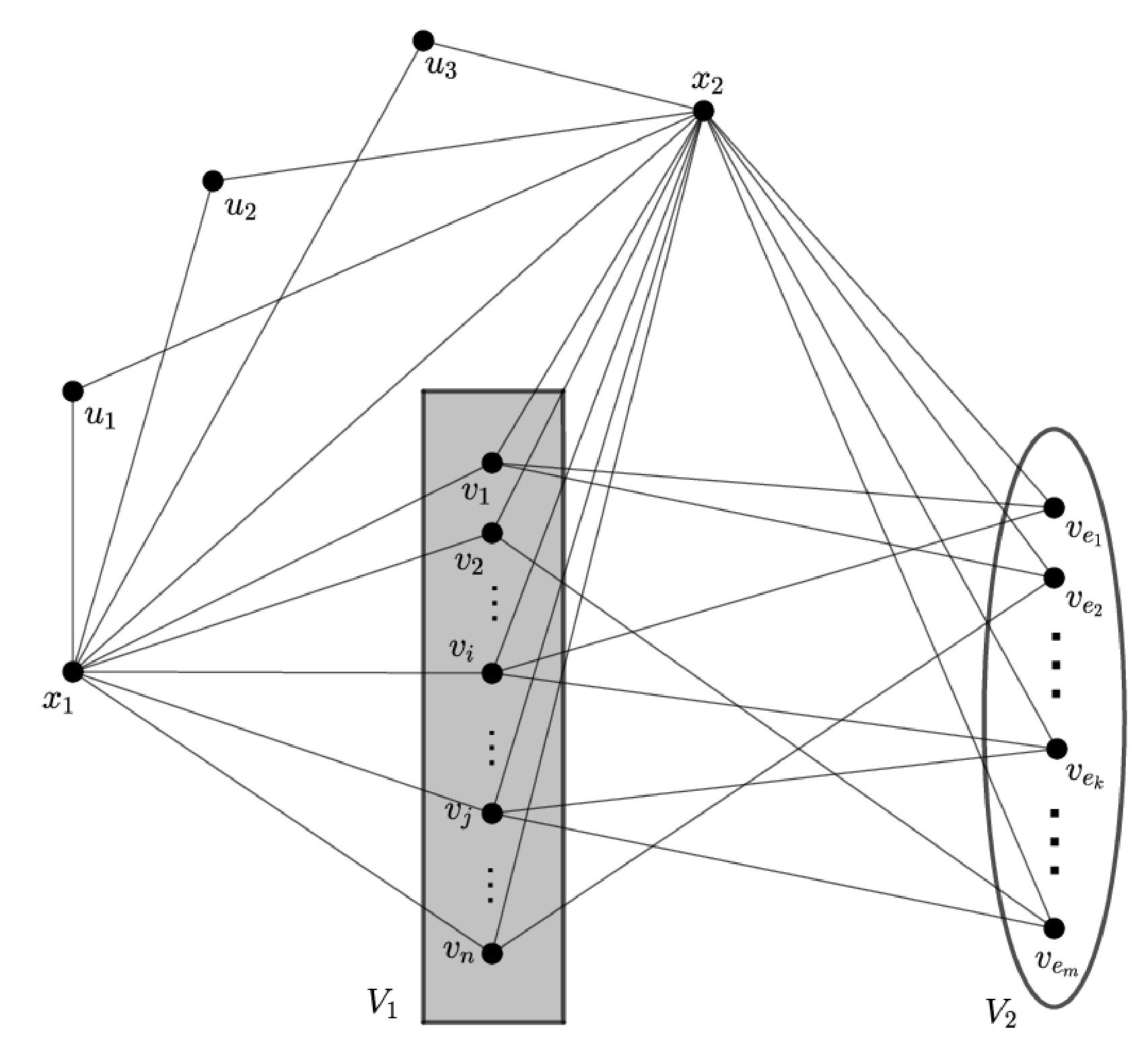}
				\caption{Construction of the graph $G'$ from $G$.
					The gray rectangle represents a clique.}
				\label{np_split_figure}
			\end{center}	
		\end{figure}
		
		Note that $G'$ is a split graph with $V(G)=C\cup I$ where $C = V_1\cup\{x_1, x_2\}$ $I = V_2\cup\{v_1, v_2, v_3\}$.
		
		\begin{claim}
			$G$ has a vertex cover of cardinality $k$ if and only if $G'$ has a dual-server dominating set of cardinality $k+2$.
		\end{claim}
		\begin{proof}

			Let $G$ has a vertex cover $S$ of cardinality $k$. Let $D= R\cup B$ such that $R=S\cup\{x_1\}$ and $B = \{x_2\}$. It is easy to see that $|D| = |S|+2 = k+2$. Now, we prove that $D$ is a dual-server dominating set. Since $S$ is a vertex cover of $G$, every vertex of $V_2$ has a neighbor in $R$. Since every vertex of $V_2$ is adjacent to $x_2$, every vertex of $V_2$ has a neighbor in $R$ and a neighbor in $B$. Since every vertex of $V_1$ is adjacent to $x_1$ and $x_2$, every vertex of $V_1$ has a neighbor in $R$ and a neighbor in $B$. Each vertex of the set $v_1, v_2, v_3$ has a neighbor in $R$ and a neighbor in $B$. Thus, $D$ is a dual-server dominating set.

			\medskip

			Let $G'$ have a dual-server dominating set $D$ of cardinality $k+2$. Note that each of $u_1$, $u_2$, and $u_3$ has exactly two neighbors, namely $x_1$ and $x_2$, in $G'$. At least two vertices from the set $\{u_1, u_2, u_3, x_1, x_2\}$ must be in $D$. Therefore, $|D\cap (V_1\cup V_2)|\leq k$. Let $S=D\cap (V_1\cup V_2)$. Since $D$ is a dual-server dominating set of $G'$ and each vertex of the set $V_2$ is adjacent to at most one vertex outside $V_1$, each vertex of $V_2\setminus S$ must have a neighbor in $V_1$. For each vertex $v\in S \cap V_2$ with $N(v)\cap S = \emptyset$, choose a neighbor $v' \in N(v) \cap V_1$ ($v'$ exists by construction) and update $S := (S \setminus \{v\}) \cup \{v'\}$. Since vertices of $V_2$ correspond to edges in $G$, $S$ is a vertex cover of $G$ with cardinality at most $k$.
		\end{proof}

		Therefore, \textsc{DualServer} is NP-complete in split graphs.
	\end{proof}

	\section{Hardness of approximation}
	\label{sec:hardness}

	In this section, we prove that, unless P=NP, the dual-server domination number cannot be approximated in polynomial time within a factor of $n^{1-\varepsilon}$ for any fixed $0<\varepsilon<1$, where $n$ is the order of the input graph. This result holds even for bipartite graphs.
	We use a gap reduction from the well-known \textsf{NP}-complete problem \textsc{3-SAT} (see \cite{cook71,GJ}).
	
	\medskip
	\begin{center}
		\noindent\fbox{\parbox{160mm}{
				\noindent \underline{\textsc{3-SAT:}}\\
				\textbf{Instance:} A collection $\mathcal C=\{C_1,C_2,\ldots,C_p\}$ of clauses on a nonempty set $X=\{x_1,x_2,\ldots,x_t\}$ of Boolean variables such that $|C_j|=3$ for each $j\in[p]$.\\
				\textbf{Question:} Is there a truth assignment for $X$ that satisfies all the clauses in $\mathcal C$?}}
	\end{center}
	
	Let $(\mathcal C,X)$ be an instance of \textsc{3-SAT}, where
	$X=\{x_1,x_2,\ldots,x_t\}$, $\mathcal C=\{C_1,C_2,\ldots,C_p\}$, and $C_j=(\ell_{j,1}\vee\ell_{j,2}\vee\ell_{j,3})$ for each $j\in[p]$. Let $\varphi$ be a truth assignment for $X$, and let $\ell_{j,s}$ be a literal of $C_j$. If $\ell_{j,s}=x_i$ for some $i\in[t]$, then $\ell_{j,s}$ is a \emph{true literal} under $\varphi$ if $\varphi(x_i)=1$. If $\ell_{j,s}=\overline{x_i}$ for some $i\in[t]$, then $\ell_{j,s}$ is a true literal under $\varphi$ if $\varphi(x_i)=0$.
	
	To establish Lemmas~\ref{lem:yes} and~\ref{lem:no}, we construct a family of graphs $G_{(\mathcal C,X),r}$, each associated with an instance $(\mathcal C,X)$ of \textsc{3-SAT} and an integer $r\ge1$. The construction is described below. We use the same symbol for a literal and the vertex representing it. In particular, $x_i$ and $\overline{x_i}$ denote two distinct vertices.
	
	\medskip
	\noindent\textbf{Construction of $G_{(\mathcal C,X),r}$:}
	We describe the construction in four steps: clause gadgets, variable gadgets, an additional structure, and the edges connecting these components.
	
	\medskip
	\noindent\textit{Clause gadgets:} For each $j\in[p]$, introduce an independent set $Z_j=\{z_{j,s}:s\in[2r(t+1)+1]\}$.
	
	\medskip
	\noindent\textit{Variable gadgets:} For each $i\in[t]$, introduce two vertices $x_i,\overline{x_i}$ and an independent set $Y_i=\{y_{i,s}:s\in[2r(t+1)+1]\}$. Join every vertex of $Y_i$ to both $x_i$ and $\overline{x_i}$.
	
	\medskip
	\noindent\textit{Additional structure:} Introduce two vertices $a,b$ and an independent set $A=\{a_s:s\in[2r(t+1)+1]\}$. Join every vertex of $A$ to both $a$ and $b$.
	
	\medskip
	\noindent\textit{The graph $G_{(\mathcal C,X),r}$:} For each $j\in[p]$, join every vertex of $Z_j$ to $a$ and to the three vertices $\ell_{j,1},\ell_{j,2},\ell_{j,3}$.
	
\begin{figure}[h]
		\begin{center}
			\includegraphics[width=0.5\linewidth]{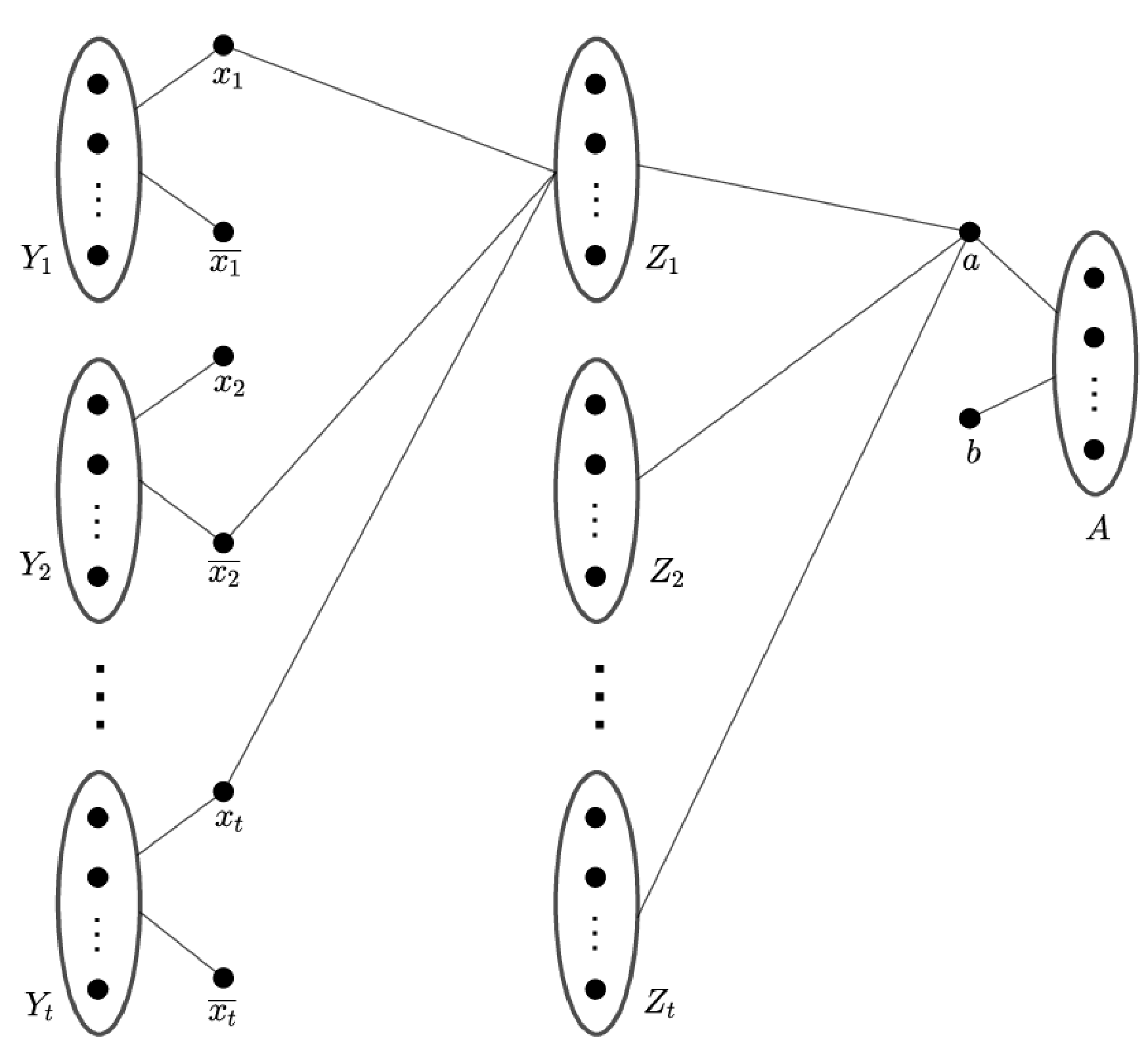}
			\caption{Construction of the graph $G_{(\mathcal C,X),r}$. A line joining a vertex to a set indicates that the vertex
				is adjacent to every vertex of that set.}
			\label{fig:construction}
		\end{center}	
	\end{figure}

	Note that the constructed graph $G_{(\mathcal C,X),r}$ is bipartite, with partite sets
	$\{a,b\}\cup\{x_i,\overline{x_i}:i\in[t]\}$ and
	$A\cup\bigcup_{i\in[t]}Y_i\cup\bigcup_{j\in[p]}Z_j$.
	The graph $G_{(\mathcal C,X),r}$ for $X=\{x_1,x_2,x_3\}$ and $\mathcal C=\{C_1,C_2,C_3\}$ is illustrated in Figure~\ref{fig:construction}. For clarity, among the edges joining clause gadgets to literal vertices, only the edges between $Z_1$ and the literal vertices of $C_1$ are shown, where $C_1=(x_1\vee\overline{x_2}\vee x_t)$.
	
	\begin{lemma}\label{lem:yes}
		Let $(\mathcal C,X)$ be an instance of \textsc{3-SAT} with $t$ variables and $p$ clauses. For every integer $r\ge 1$, if $\mathcal C$ is satisfiable, then $\gds(G_{(\mathcal C,X),r})\le 2(t+1)$.
	\end{lemma}
	
	\begin{proof}
		Let $G=G_{(\mathcal C,X),r}$, and let $\varphi$ be a truth assignment for $X$ that satisfies every clause in $\mathcal C$. Let $S=\{a,b\}\cup\{x_i,\overline{x_i}:i\in[t]\}$. Define a partition $\{R,B\}$ of $S$ as follows. Let $a\in R$ and $b\in B$. For each $i\in[t]$, let $x_i\in B$ and $\overline{x_i}\in R$ if $\varphi(x_i)=1$, and let $x_i\in R$ and $\overline{x_i}\in B$ if $\varphi(x_i)=0$. Thus, for every $j\in[p]$ and $k\in[3]$, $\ell_{j,k}\in B$ if and only if $\ell_{j,k}$ is a true literal under $\varphi$.
		
		We show that every vertex in $V(G)\setminus S$ has a neighbor in $R$ and a neighbor in $B$. By construction, every vertex of $A$ is adjacent to $a\in R$ and $b\in B$. For each $i\in[t]$, every vertex of $Y_i$ is adjacent to both $x_i$ and $\overline{x_i}$, one of which lies in $R$ and the other in $B$. For each $j\in[p]$, every vertex of $Z_j$ is adjacent to $a\in R$ and to $\ell_{j,1},\ell_{j,2},\ell_{j,3}$. Since $\varphi$ satisfies $C_j$, at least one of these literals is true under $\varphi$, and its corresponding vertex lies in $B$. Therefore, $S$ is a dual-server dominating set of $G$ and hence $\gds(G)\le |S|=2(t+1)$.
	\end{proof}

	\begin{lemma}\label{lem:no}
		Let $(\mathcal C,X)$ be an instance of \textsc{3-SAT} with $t$ variables and $p$ clauses. For every integer $r\ge 1$, if $\mathcal C$ is unsatisfiable, then $\gds(G_{(\mathcal C,X),r})>2r(t+1)$.
	\end{lemma}
	
	\begin{proof}
		Let $G=G_{(\mathcal C,X),r}$. Suppose, to the contrary, that $\gds(G)\le 2r(t+1)$. Let $S$ be a minimum dual-server dominating set of $G$, and let $\{R,B\}$ be a partition of $S$ such that every vertex in $V(G)\setminus S$ has a neighbor in $R$ and a neighbor in $B$. Since $|S|\le 2r(t+1)$ and each of the sets $A,Y_1,\ldots,Y_t,Z_1,\ldots,Z_p$ has $2r(t+1)+1$ vertices, each contains at least one vertex outside $S$.
		
		\begin{claim}\label{cl:no-ab}
			One of $a,b$ belongs to $R$ and the other belongs to $B$.
		\end{claim}
		\begin{proof}[Proof of Claim~\ref{cl:no-ab}]
			Choose $a_s\in A\setminus S$. By construction, $N_G(a_s)=\{a,b\}$. Since $a_s$ has a neighbor in $R$ and a neighbor in $B$, and $R\cap B=\emptyset$, the claim follows.
		\end{proof}
		
		By Claim~\ref{cl:no-ab}, interchanging $R$ and $B$ if necessary, we may assume that $a\in R$ and $b\in B$.
		
		\begin{claim}\label{cl:no-var}
			For every $i\in[t]$, one of $x_i,\overline{x_i}$ belongs to $R$ and the other belongs to $B$.
		\end{claim}
		\begin{proof}[Proof of Claim~\ref{cl:no-var}]
			Fix $i\in[t]$, and choose $y_{i,s}\in Y_i\setminus S$. By construction, $N_G(y_{i,s})=\{x_i,\overline{x_i}\}$. Since $y_{i,s}$ has a neighbor in each of the disjoint sets $R$ and $B$, one of $x_i,\overline{x_i}$ belongs to $R$ and the other belongs to $B$.
		\end{proof}
		
		Define a truth assignment $\varphi$ for $X$ by setting $\varphi(x_i)=1$ if $x_i\in B$, and $\varphi(x_i)=0$ otherwise. By Claim~\ref{cl:no-var}, $\overline{x_i}\in B$ if and only if $\varphi(x_i)=0$. Thus, for every $j\in[p]$ and $k\in[3]$, $\ell_{j,k}\in B$ if and only if $\ell_{j,k}$ is a true literal under $\varphi$.
		
		\begin{claim}\label{cl:no-sat}
			The assignment $\varphi$ satisfies every clause in $\mathcal C$.
		\end{claim}
		\begin{proof}[Proof of Claim~\ref{cl:no-sat}]
			Fix $j\in[p]$, and choose $z_{j,s}\in Z_j\setminus S$. By construction, $N_G(z_{j,s})=\{a,\ell_{j,1},\ell_{j,2},\ell_{j,3}\}$. Since $a\in R$ and $z_{j,s}$ has a neighbor in $B$, at least one of $\ell_{j,1},\ell_{j,2},\ell_{j,3}$ belongs to $B$. The corresponding literal is true under $\varphi$, so $\varphi$ satisfies $C_j$.
		\end{proof}
		
		By Claim~\ref{cl:no-sat}, $\mathcal C$ is satisfiable, contradicting the hypothesis. Therefore, $\gds(G)>2r(t+1)$.
	\end{proof}
	
	By Lemmas~\ref{lem:yes} and~\ref{lem:no}, we have the following corollary.

	\begin{coro}\label{cor:gap}
		Let $(\mathcal C,X)$ be an instance of \textsc{3-SAT} with $t$ variables and $p$ clauses, and let $r\ge 1$ be an integer. Then, the following holds

		\[\gds(G_{(\mathcal C,X),r})\begin{cases} 
			\le 2(t+1),~\text{if $C$ is satisfiable,}\\
			> 2r(t+1),~\text{if $C$ is not satisfiable.}\\
			
		\end{cases}\]
	\end{coro}

	\begin{theorem}\label{thm:bipartite}
		Unless P=NP, the dual-server domination number cannot be approximated in polynomial time within a factor of $n^{1-\varepsilon}$ for any fixed $0<\varepsilon<1$, even for bipartite graphs, where $n$ is the order of the input graph.
	\end{theorem}
	
	\begin{proof}
		Let $0<\varepsilon<1$ be fixed, and let $(\mathcal C,X)$ be an instance of \textsc{3-SAT} with $t$ variables and $p$ clauses. We set $s=3t+p+3$,$q=\lceil 2/\varepsilon\rceil$, and $ r=s^q$.
		Let $G=G_{(\mathcal C,X),r}$ be the bipartite graph associated with $(\mathcal C,X)$ and $r$ as described by the preceding construction. Moreover, let $n=|V(G)|$. By construction,
		\[
		n=2(t+1)+(t+p+1)\bigl(2r(t+1)+1\bigr)
		=s+2r(t+1)(t+p+1).
		\]
		Note that the graph $G$ can be constructed in polynomial time. We first establish the following claim.
		
		\begin{claim}\label{cl:inapprox-r}
			Unless P=NP, there is no polynomial-time algorithm that approximates $\gds(G_{(\mathcal C,X),r})$ within a factor of $r=s^q$.
		\end{claim}
		
		\begin{proof}[Proof of Claim~\ref{cl:inapprox-r}]
			Suppose that $\mathcal Appr$ is a polynomial-time algorithm that achieves an approximation factor of $r$ on the constructed graphs. We show that $\mathcal Appr$ can be used to solve \textsc{3-SAT} in polynomial time. Let $S$ be the dual-server dominating set returned by $\mathcal Appr$ when applied to $G$. Then $\gds(G)\le |S|\le r\cdot\gds(G)$.
			If $\mathcal C$ is satisfiable, then Corollary~\ref{cor:gap} yields
			\[
			|S|\le r\gds(G)\le 2r(t+1).
			\]
			On the other hand, if $\mathcal C$ is not satisfiable, then Corollary~\ref{cor:gap} gives
			\[
			|S|\ge \gds(G)>2r(t+1).
			\]
			Thus, by comparing $|S|$ with $2r(t+1)$, we can determine in polynomial time whether $\mathcal C$ is satisfiable. Since \textsc{3-SAT} is NP-complete, this implies P=NP.
		\end{proof}
		
		We now relate $r$ to the order $n$ of $G$. Since
		$2(t+1)\le s-1$, $t+p+1\le s-1$, and $r\ge1$, we obtain
		\begin{equation}\label{eq:order-bound}
			n=s+2r(t+1)(t+p+1)
			\le r\bigl(s+(s-1)^2\bigr)
			\le rs^2
			=s^{q+2}.
		\end{equation}
		By the choice of $q$, we have $\varepsilon q\ge2$, and hence
		\[
		(q+2)(1-\varepsilon)
		=q+2-\varepsilon q-2\varepsilon
		\le q-2\varepsilon<q.
		\]
		Since $s>1$, it follows from \eqref{eq:order-bound} that
		\[
		n^{1-\varepsilon}
		\le s^{(q+2)(1-\varepsilon)}
		<s^q=r.
		\]
		Consequently, a polynomial-time approximation algorithm with factor
		$n^{1-\varepsilon}$ for bipartite graphs would also achieve a factor
		of $r$ on the constructed graphs, contradicting
		Claim~\ref{cl:inapprox-r} unless P=NP.
		This completes the proof.
	\end{proof}

	\section{Dual-server domination in threshold graphs}
	
	A graph $G$ is a \emph{threshold graph} if there exist non-negative real numbers $w_v$ for each $v\in V(G)$ and $t$ such that $w(U)<t$ if and only if $U\subseteq V(G)$ is a stable set, where $w(U)=\sum_{v\in U}w_v$. Equivalently, a graph $G$ is a threshold graph if it can be constructed from the one-vertex graph by repeatedly adding either an isolated vertex or a dominating vertex. There are a few more equivalent definitions of threshold graphs; see~\cite{Mahadev1995}. By Theorem~\ref{split:np_complete}, \textsc{DualServer} is NP-complete on split graphs. In this section, we study the dual-server domination number for threshold graphs, which is a proper subclass of split graphs. First, we prove that $\gamma_{2}(G)=\gds(G)$ for each threshold graph $G$. Consequently, a polynomial time algorithm exists to compute $\gds(G)$ for each threshold graph $G$; see~\cite{CMV2013}. We completely determine the dual-server domination number of a threshold graph $G$, which in turn can be used to compute the dual-server domination number of $G$ in linear time. We use the following characterization of threshold graphs, which is used in designing efficient algorithms. 
	
	\begin{propo}[\cite{Mahadev1995}]
		A graph $G$ is a threshold graph if and only if $G$ is a split graph with split partition $(\mathcal{C},\mathcal{I})$ with $\mathcal{C}=\{c=x_1, x_2, \ldots, x_p\}$ and $\mathcal{I}=\{y_1, y_2, \ldots, y_q\}$ such that $N[x_1]\subseteq N[x_2]\subseteq\ldots\subseteq N[x_p]$ and $N(y_1)\supseteq N(y_2)\supseteq\ldots\supseteq N(y_q)$.
	\end{propo}
	
	First, we prove that $\gamma_2(G)=\gamma_{{ds}}(G)$ for a threshold graph $G$.

	\begin{propo}
		If $G$ is a threshold graph, then
		$\gamma_2(G)=\gamma_{ds}(G)$.
	\end{propo}
	
	\begin{proof}
		By Observation~\ref{prop:twodom}, $\gamma_2(G)\le\gamma_{{ds}}(G)$. It remains to show that $\gamma_{ds}(G)\leq\gamma_2(G)$.
		We show that every $2$-dominating set of $G$ is a dual-server dominating set. Let $S$ be a $2$-dominating set of $G$. If $S=V(G)$, then $S$ is trivially a dual-server dominating set.
		Hence, we assume that $S\neq V(G)$.
		
		\medskip
		
		\noindent\textbf{Case 1.} $S\cap C\neq\emptyset$.
		
		\medskip
		
		Let $x_i$ be the largest indexed vertex in $S\cap C$. Let $R=\{x_i\}$ and $B=S\setminus R$. We show that every vertex in $V(G)\setminus S$ has a neighbor in $R$ and a neighbor in $B$. Let $v\in C\setminus S$. Since $C$ is a clique,
		$v$ is adjacent to $x_i$. Moreover, since $S$ is a
		$2$-dominating set, $v$ has at least one more neighbor
		in $S$, which belongs to $B$. Now, let $v\in I\setminus S$. Since $I$ is an independent set and $S$ is a $2$-dominating set, $v$ has at least two neighbors in $S\cap C$. Since $N[x_1]\subseteq N[x_2]\subseteq\cdots\subseteq N[x_p]$ and $x_i$ is the largest indexed vertex in $S\cap C$,
		$v$ is adjacent to $x_i$. Since $v$ has at least two
		neighbors in $S$, it also has a neighbor in $B$. Therefore, every vertex in $V(G)\setminus S$ has
		a neighbor in both $R$ and $B$.

		\medskip
		
		\noindent\textbf{Case 2.} $S\cap C=\emptyset$.
		
		\medskip
		
		Since $I$ is an independent set and $S$ is a $2$-dominating set, we have $I\subseteq S$. Consequently, $S=I$. Let $R=\{y_1\}$ and $B=S\setminus R$. Consider any vertex $v\in V(G)\setminus S$. Since $S$ is a $2$-dominating set, $v$ has at least two neighbors in $S$. Since $N(y_1)\supseteq N(y_2)\supseteq\cdots\supseteq N(y_q)$, $v$ is adjacent to $y_1$. Moreover, $v$ has at least one more neighbor in $S$, which belongs to $B$. Thus, every vertex in $V(G)\setminus S$ has a neighbor in both $R$ and $B$.
		
		In both cases, $S$ is a dual-server dominating set. Consequently, $\gamma_{ds}(G)\leq\gamma_2(G)$.
		Hence, $\gamma_2(G)=\gamma_{ds}(G)$.
	\end{proof}

	\begin{theorem}\label{threshold_characterization_pcf}
		Let $G$ be a threshold graph such that $V(G)=\mathcal{C}\cup\mathcal{I}$ and $\mathcal{C}\cap\mathcal{I}=\emptyset$, where $\mathcal{C}$ is a clique and $\mathcal{I}$ is an independent set. Let $\mathcal{C}=\{x_1, x_2, \ldots, x_p\}$ and $\mathcal{I}=\{y_1, y_2, \ldots, y_q\}$ be such that $N[x_1]\subseteq N[x_2]\subseteq\ldots\subseteq N[x_p]$ and $N(y_1)\supseteq N(y_2)\supseteq\ldots\supseteq N(y_q)$. Then,
		
		$$\gamma_{ds}(G)=
		\begin{cases}
			n,
			& \text{if } n\le2,\\[1mm]
			t,
			& \text{if } G\cong K_{1,t}
			\text{ and } t\ge2,\\[1mm]
			2+\left|N[x_p]\setminus N[x_{p-1}]\right|,
			& \text{otherwise}.
		\end{cases}$$

	\end{theorem}
	\begin{proof}
		By Observation~\ref{pendant_observation_ds}, every vertex of degree at most one belongs to every dual-server
		dominating set. If $n\le2$, every vertex of $G$ has degree at most one. Consequently, every vertex must be in the dual-server dominating set. Consequently, $\gamma_{ds}(G)=n$ when $n\le2$.
		
		Now, assume that $G\cong K_{1,t}$ with $t\ge2$. Let $V(G)$ is partitioned into two sets $X$ and $Y$ such that $X=\{x\}$ and $Y=\{y_1, y_2, \ldots, y_t\}$ and $X$ and $Y$ both are independent sets. Note that each vertex in $Y$ is a pendant vertex. Therefore, $\gamma_{ds}(G)\ge t$, by Observation~\ref{pendant_observation_ds}. Conversely, let $S=R\cup B$ be such that $R=Y\setminus\{v_1\}$ and $B=\{v_1\}$. Since $t\ge2$, both $R$ and $B$ are nonempty. 
		Note that $|S|=t$ and each vertex of $V(G)\setminus S$ has a neighbor in $R$ and a neighbor in $B$. Therefore, $S$ is a dual-server dominating set with cardinality at most $t$. Consequently, $\gamma_{ds}(G)=n-1$.

		Finally, assume that $G$ is not a star and has at least three vertices. Since $G$ is connected and is not a star, we have $p\ge2$. Let $P$ denote the set of pendant vertices of $G$. We first show that $P=N[x_p]\setminus N[x_{p-1}]$. It is easy to see that $N[x_p]=V(G)$. Since $G$ is not a star, no vertex of $\mathcal{C}$ is a pendant vertex. Therefore, $P\subseteq\mathcal I$. Let $y\in N[x_p]\setminus N[x_{p-1}]$. 
		Since $N[x_1]\subseteq N[x_2]\subseteq\ldots\subseteq N[x_p]$, $y$ is not adjacent to $x_i$ for each $i\le p-1$. Since $\mathcal I$ is independent, we have $N(y)=\{x_p\}$. Consequently, $y\in P$. Conversely, let $y\in P$.
		Since $y\in\mathcal {I}$ and $N[x_p]=V(G)$, we have$N(y)=\{x_p\}$. Thus, we have $P=N[x_p]\setminus N[x_{p-1}]$.
		
		We now establish the upper bound. Let $R=\{x_p\}$, $B=P\cup\{x_{p-1}\}$, and $D=R\cup B$. Note that neither $x_p$ nor $x_{p-1}$ belongs to $P$. Consequently, $R$ and $B$
		are disjoint and $|D|=|P|+2$. Consider any vertex $z\in V(G)\setminus D$. Since $N[x_p]=V(G)$, $z$ is adjacent to $x_p$. Since $P=N[x_p]\setminus N[x_{p-1}]$ and $N[x_p]=V(G)$, we have $V(G)\setminus P=N[x_{p-1}]$. As $z\notin P$ and $z\ne x_{p-1}$, the vertex $z$ is
		adjacent to $x_{p-1}$. Therefore $z$ has a neighbor in both $R$ and $B$. Consequently, $\gamma_{ds}(G)\le |P|+2$.
		
		Now, we show that $\gamma_{ds}(G)\ge |P|+2$. For the sake of contradiction, assume that $D'$ is a dual-server dominating set with $|D'|\le |P|+1$. By Observation~\ref{pendant_observation_ds}, we have $P\subseteq D'$. Therefore, $|D'\setminus P|\le1$. Since $x_{p-1}$ is not a pendant vertex and is adjacent to $x_p$, it has a neighbor $w\ne x_p$. Consequently, $G$ has at least three vertices having degree at least $2$, namely $x_p$, $x_{p-1}$, and $w$. Since $D'$ contains at most one vertex from $V(G)\setminus P$, at least one of the vertices $x_{p-1}$ and $w$ does not belong to $D'$. Let $z\in\{x_{p-1}, w\}$ be a vertex such that $z\notin D'$. Since $z\ne x_p$, we obtain $N(z)\cap P=\emptyset$. Therefore, $|N(z)\cap D'|=|N(z)\cap(D'\setminus P)|\le1$. Since $z\notin D'$ and has at most one neighbor in $D'$, we have $D'$ is not a dual-server dominating set. Therefore, $\gamma_{ds}(G)\ge |P|+2$.

		Hence, $\gamma_{ds}(G)=|P|+2 =2+\left|N[x_p]\setminus N[x_{p-1}]\right|$. This completes the proof.
	\end{proof}

	\section{Dual-server domination in chain graphs}
	\label{sec:chain-graphs}

	A bipartite graph $G= (X \cup Y, E)$ is a \emph{chain graph} if there exists orderings $\sigma_X =\{x_1, x_2, \ldots, x_p\}$ and $\sigma_Y=\{y_1, y_2, \ldots, y_q\}$ such that $N(x_1)\subseteq N(x_2)\subseteq\ldots\subseteq N(x_p)$ and $N(y_1)\supseteq N(y_2)\supseteq\ldots\supseteq N(y_q)$. In this section, we first show that $\gamma_2(G)=\gds(G)$ for every chain graph $G$. By Theorem~\ref{Dserver:np_complete}, \textsc{DualServer} is NP-complete on chordal bipartite graphs. In this section, we study the dual-server domination number for chain graphs, which is a proper subclass of chordal bipartite graphs. We completely determine the dual-server domination number of a chain graph $G$, which in turn can be used to compute the dual-server domination number of $G$ in linear time.

	\begin{lemma}\label{lem:chain-assignment}
		Every $2$-dominating set $D$ of a chain graph $G$ admits a partition $D=R\cup B$ such that $|R\cap X|\le 1$ and $|R\cap Y|\le 1$ and every vertex $v\in V(G)\setminus D$ has a neighbor in $R$ and a neighbor in $B$.
		
	\end{lemma}
	
	\begin{proof}
		If $D=V(G)$, any assignment is valid; assign at most one vertex
		to $R$ and all remaining vertices to $B$. So we assume that
		$D\ne V(G)$. If $D\cap X\ne\emptyset$, then choose a maximum-degree vertex
		$x^*\in D\cap X$. Similarly, choose a maximum-degree vertex
		$y^*\in D\cap Y$ if $D\cap Y\ne\emptyset$. Let $R=\{x^*, y^*\}$ and $B=D\setminus R$. Let $z\in Y\setminus D$. Since $D$ is $2$-dominating, $z$
		has two distinct neighbors in $D\cap X$. By the nested-neighborhood property of chain graphs, $z$ is adjacent to $x^*$. At least one other neighbor
		of $z$ in $D\cap X$ belongs to $B$. Thus, $z$ has a neighbor in $R$ and a neighbor in $B$. Let $z\in X\setminus D$. Since $D$ is $2$-dominating, $z$
		has two distinct neighbors in $D\cap Y$. By the nested-neighborhood property of chain graphs, $z$ is adjacent to $y^*$. At least one other neighbor of $z$ in $D\cap Y$ belongs to $B$. Thus, $z$ has a neighbor in $R$ and a neighbor in $B$. Hence, every $2$-dominating set $D$ of a chain graph $G$ admits a partition $D=R\cup B$ such that $|R\cap X|\le 1$ and $|R\cap Y|\le 1$ and every vertex $v\in V(G)\setminus D$ has a neighbor in $R$ and a neighbor in $B$.
	\end{proof}
	
	\begin{coro}\label{cor:chain-equality}
		For every chain graph $G$, $\gamma_{ds}(G)=\gamma_2(G)$.
	\end{coro}
	
	\begin{proof}
		By Observation~\ref{prop:twodom}, $\gamma_{ds}(G)\ge\gamma_2(G)$. By Lemma~\ref{lem:chain-assignment}, every $2$-dominating set of $G$ is a dual-server dominating set. Therefore, $\gamma_{ds}(G)\le\gamma_2(G)$. Hence, $\gamma_{ds}(G)=\gamma_2(G)$ for every chain graph $G$.
	\end{proof}
	
	Now, we determine the dual-server domination number for chain graphs. The graph $K_1$ has $\gamma_{ds}(K_1)=\gamma_2(K_1)=1$, so we assume that $G$
	is connected and has at least two vertices.
	Let $P=\{z\in V(G):d(z)=1\}$, $X_0=X\setminus P$, $Y_0=Y\setminus P$, and set $\ell=|P|$, $a=|X_0|$, $b=|Y_0|$. By observation~\ref{pendant_observation_ds}, every vertex of $P$ belongs to every dual-server dominating set. We define
	
	$$\eta_X=
	\begin{cases}
		1,
		& \text{if $|P\cap Y|\ge2$ and $d(y)\ge3$
			for every $y\in Y_0$,}\\
		0,
		& \text{otherwise,}
	\end{cases}$$
	
	and
	
	$$\eta_Y=
	\begin{cases}
		1,
		& \text{if $|P\cap X|\ge2$ and $d(x)\ge3$
			for every $x\in X_0$,}\\
		0,
		& \text{otherwise.}
	\end{cases}$$

	Equivalently, $\eta_X=1$ exactly when $Y$ contains at least two
	pendant vertices and no vertex of degree $2$; the condition for
	$\eta_Y$ is symmetric.
	
	\begin{theorem}\label{thm:chain-formula}
		Let $G=(X\cup Y,E)$ be a connected chain graph with at least two
		vertices. With the notation above,  
		$\gamma_{ds}(G)=\ell+\min\{a-\eta_X,\ b-\eta_Y,\ 4\}$.
	\end{theorem}
	
	\begin{proof}
		By Corollary~\ref{cor:chain-equality}, it suffices to determine
		$\gamma_2(G)$. Let $u\in X$ and $v\in Y$ be vertices of maximum
		degree in $X$ and $Y$, respectively. Since $G$ is a connected
		chain graph, we have $N(u)=Y$ and $N(v)=X$. Consequently,
		every pendant vertex in $Y$ is adjacent only to $u$, and
		every pendant vertex in $X$ is adjacent only to $v$.
		By Observation~\ref{pendant_observation_ds}, $P$ is contained in every $2$-dominating set of $G$.
		
		\medskip
		\noindent\textbf{Upper bound.}
		First, we show that $\gamma_2(G)\le \ell+\min\{a-\eta_X,b-\eta_Y,4\}$. Let $D_X=P\cup X_0$. Since $X\subseteq D_X$, every vertex in $V(G)\setminus D_X$ belongs to $Y_0$ and has at least two neighbors in $D_X$. Therefore, $D_X$ is a $2$-dominating set of $G$ of cardinality $\ell+a$. Assume that $\eta_X=1$. By the definition of $\eta_X$,
		the vertex $u$ has at least two pendant neighbors, and
		$d(y)\ge3$ for every vertex $y\in Y_0$. We claim that
		$D_X\setminus\{u\}$ is also a $2$-dominating set of $G$.
		Indeed, $u$ has at least two neighbors in $P$, and every
		vertex $y\in Y_0$ has at least two neighbors in
		$X\setminus\{u\}$. Thus, every vertex $z\in V(G)\setminus (D_X\setminus\{u\})$ has at least two neighbors in $D_X\setminus\{u\}$. Consequently, $\gamma_2(G)\le \ell+a-1$.
		Hence, we have $\gamma_2(G)\le \ell+a-\eta_X$. By the similar argument, we have $\gamma_2(G)\le \ell+b-\eta_Y$.

		It remains to show that $\gamma_2(G)\le\ell+4$.
		If $a\le1$ or $b\le1$, this follows immediately from
		the previous two inequalities. Therefore, assume that
		$a,b\ge2$. Let $u_1,u_2$ be two vertices of largest degree
		in $X_0$, and let $v_1,v_2$ be two vertices of largest degree
		in $Y_0$. Consider the set $D=P\cup\{u_1,u_2,v_1,v_2\}$. Let $x\in X\setminus D$. Since $G$ is a chain
		graph, $x$ is adjacent to both $v_1$ and $v_2$.
		Similarly, every vertex in $Y\setminus D$ is adjacent to
		both $u_1$ and $u_2$. Therefore, $D$ is a $2$-dominating
		set of $G$ of cardinality $\ell+4$.
		
		Combining the three upper bounds, we obtain $\gamma_2(G)\le \ell+\min\{a-\eta_X,b-\eta_Y,4\}$.

		\medskip
		\noindent\textbf{Lower bound.}
		Let $D$ be a minimum $2$-dominating set of $G$. Since $P\subseteq D$, we have $|D|=\ell+|D\cap X_0|+|D\cap Y_0|$.

		\medskip
		\noindent\textbf{Case 1.}
		At least one of $D\cap X_0$ and $D\cap Y_0$ is empty.
		\medskip
		
		Without loss of generality, assume that $D\cap Y_0=\emptyset$.
		Let $x\in X_0\setminus\{u\}$. Since $u$ is the only vertex
		in $X$ that can have pendant neighbors, we have
		$N(x)\cap P=\emptyset$. Moreover, since
		$D\cap Y_0=\emptyset$, the vertex $x$ has no neighbor in $D$.
		Therefore, $x$ must belong to $D$. Consequently, $X_0\setminus\{u\}\subseteq D$. If $X_0\subseteq D$, then $|D|\ge\ell+a\ge\ell+a-\eta_X$. So, we assume that $X_0\not\subseteq D$. Since $D$ is a $2$-dominating set and $u\in X$ is the only vertex in $X$ that has neighbors in $P$, we have $u\in X_0\setminus D$ and $D=P\cup(X_0\setminus\{u\})$. Since $u\notin D$ and $D$ is a $2$-dominating set, $u$ must have at least two neighbors in $P\cap Y$. Thus, $|P\cap Y|\ge2$. Furthermore, let $y\in Y_0$. Since $D\cap Y_0=\emptyset$, we have $y\notin D$. Since $u$ is adjacent to $y$ and every vertex of $X$ except $u$ belongs to $D$, we have $|N(y)\cap D|=d(y)-1$. As $D$ is a $2$-dominating set, it follows that $d(y)\ge3$. Consequently, we have $d(y)\ge3$ for every $y\in Y_0$. Therefore, $\eta_X=1$, and hence $|D|=\ell+a-1=\ell+a-\eta_X$. Thus, in either case, $|D|\ge\ell+a-\eta_X$.
		
		\medskip
		\noindent\textbf{Case 2.}
		Both $D\cap X_0$ and $D\cap Y_0$ are nonempty, and at
		least one of them contains exactly one vertex.
		
		\medskip
		
		Without loss of generality, assume that $|D\cap X_0|=1$.
		Consider a vertex $y\in Y_0\setminus\{v\}$. Since $v$ is
		the only vertex in $Y$ that can have pendant neighbors,
		we have $N(y)\cap P=\emptyset$. Thus, $y$ has at most one
		neighbor in $D$. Since $D$ is a $2$-dominating set, we
		must have $y\in D$. Therefore, $Y_0\setminus\{v\}\subseteq D$.
		Consequently, $|D\cap Y_0|\ge b-1$, and hence $|D|=\ell+|D\cap X_0|+|D\cap Y_0|\ge\ell+1+(b-1)=\ell+b\ge\ell+b-\eta_Y$.
		
		\medskip
		\noindent\textbf{Case 3.}
		Both $D\cap X_0$ and $D\cap Y_0$ contain at least two
		vertices. 
		
		\medskip
		
		In this case, $|D|=\ell+|D\cap X_0|+|D\cap Y_0|\ge\ell+4$.
		
		It follows from the above three cases that $\gamma_2(G)=|D|\ge\ell+\min\{a-\eta_X,b-\eta_Y,4\}$.  Combining the upper and lower bounds, we obtain $\gamma_2(G)=\ell+\min\{a-\eta_X,b-\eta_Y,4\}$. Since $\gamma_{ds}(G)=\gamma_2(G)$ by
		Corollary~\ref{cor:chain-equality}, the theorem follows.
	\end{proof}

	\section{Dual-server domination in proper interval graphs}
	
	By Theorem~\ref{Dserver:np_complete}, \textsc{DualServer} is NP-complete on undirected path graphs. In this section, we study the parameter for proper interval graphs, which is a proper subclass of undirected path graphs. We show that $\gamma_2(G)=\gds(G)$ for every proper interval graph $G$. Chiarelli et al.~\cite{ChiarelliEtAl2019} proved that $\gamma_2(G)$ can be computed in $O(n^6)$ time for every proper interval graph $G$. Recently, Li et al.~\cite{LiLiShang2025} proved that $\gamma_2(G)$ can be computed in $O(n^4)$ for every proper interval graph $G$.  Consequently, $\gds(G)$ can be computed in polynomial time for every proper interval graph $G$.

	A graph $G$ is a \emph{proper interval graph} if it is the intersection graph of a family of intervals on the real line in which no interval properly contains another. A vertex $v$ of $G$ is \emph{simplicial} if $N_G[v]$ is a clique of $G$. An ordering $\sigma=(v_1,v_2,\ldots,v_n)$ of $V(G)$ is a \emph{perfect elimination ordering} (PEO) of $G$ if, for each $i\in\{1,2,\ldots,n\}$, the vertex $v_i$ is simplicial in the subgraph of $G$ induced by $\{v_i,v_{i+1},\ldots,v_n\}$. A PEO $\sigma$ of $G$ is a \emph{bicompatible elimination ordering} (BCO) if the reverse ordering $(v_n,v_{n-1},\ldots,v_1)$ is also a PEO of $G$. Note that each proper interval graph admits a BCO~\cite{Jamison1984}. Further, a BCO of a proper
	interval graph can be computed in linear time~\cite{Panda2003}. We use the
	following property of a BCO.
	
	\begin{obs}[\cite{Panda2003}]\label{obs:bco}
		Let $\sigma=(v_1,v_2,\ldots,v_n)$ be a BCO of a connected proper interval graph $G$. If $v_iv_j\in E(G)$ with $i<j$, then $\{v_i,v_{i+1},\ldots,v_j\}$ is a clique of $G$.
	\end{obs}

	\begin{theorem}\label{thm:pig}
		If $G$ is a proper interval graph, then $\gamma_{{ds}}(G)=\gamma_2(G)$.
	\end{theorem}
	
	\begin{proof}
		By Observation~\ref{prop:twodom}, $\gamma_2(G)\le\gamma_{{ds}}(G)$. It therefore suffices to prove that $\gamma_{{ds}}(G)\le\gamma_{2}(G)$. We show that every $2$-dominating set of $G$ is a dual-server dominating set of $G$. A set $D\subseteq V(G)$ is a $2$-dominating set (resp. a dual-server dominating set) of $G$ if and only if $D\cap V(H)$ is a $2$-dominating set (resp. a dual-server dominating set) of $H$ for every component $H$ of $G$. Therefore, we may assume that $G$ is connected. Let $\sigma=(v_1,v_2,\ldots,v_n)$ be a BCO of $G$. Let $D=\{v_{i_1},v_{i_2},\ldots,v_{i_m}\}$ be a $2$-dominating set of $G$ where $i_1<i_2<\cdots<i_m$. Define  $R=\{v_{i_t} : t \text{ is odd}\}$ and $B=\{v_{i_t} : t \text{ is even}\}$. Then $\{R,B\}$ is a partition of $D$. If $V(G)\setminus D=\emptyset$, then the desired result follows. So we may assume that $V(G)\setminus D\neq\emptyset$.
		
		\begin{claim}
			\label{cl:pig}
			For every vertex $v_k\in V(G)\setminus D$ there exist integers $j<\ell$ such that $N_G(v_k)\cap D=\{v_{i_j},v_{i_{j+1}},\ldots,v_{i_\ell}\}$.
		\end{claim}
		
		\begin{proof}[Proof of Claim~\ref{cl:pig}]
			Let $v_k\in V(G)\setminus D$. Since $D$ is a $2$-dominating set, $|N_G(w)\cap D|\ge 2$. Let $j$ and $\ell$ be the minimum and the maximum index $t$ such that $v_{i_t}\in N_G(v_k)\cap D$. Since $|N_G(w)\cap D|\ge 2$, we have $j<\ell$. It remains to show that $v_{i_t}\in N_G(v_k)$ for every $t$ with $j<t<\ell$. Assume, to the contrary, that $v_{i_t}v_k\notin E(G)$ for some $t$ where $j<t<\ell$. Since $v_k\notin D$, we have $i_t\neq k$. If $i_t<k$, then $i_j<i_t<k$ and $v_{i_j}v_k\in E(G)$. By Observation~\ref{obs:bco}, $\{v_{i_j},v_{i_j+1},\ldots,v_k\}$ is a clique of $G$, and hence $v_{i_t}v_k\in E(G)$, a contradiction. If $i_t>k$, then $k<i_t<i_\ell$ and $v_kv_{i_\ell}\in E(G)$. By Observation~\ref{obs:bco},
			$\{v_k,v_{k+1},\ldots,v_{i_\ell}\}$ is a clique of $G$, and again
			$v_{i_t}v_k\in E(G)$, a contradiction. Therefore, $v_{i_t}\in N_G(v_k)$ for every $t$ with $j<t<\ell$.
		\end{proof}

		Now let $v_k\in V(G)\setminus D$ and $j$ be the minimum index $t$ such that $v_{i_t}\in N_G(v_k)\cap D$ as in the Claim~\ref{cl:pig}. By Claim~\ref{cl:pig}, $v_{i_j}$ and $v_{i_{j+1}}$ are both neighbors of $v_k$. According to the construction of $R$ and $B$, one of $v_{i_j},v_{i_{j+1}}$ belongs to $R$ and the other belongs to $B$. Hence $v_k$ has a neighbor in $R$ and a neighbor in $B$. Since $v_k$ is arbitrary, $D$ is a dual-server dominating set of $G$ with associated partition $\{R,B\}$. Thus, $\gamma_{{ds}}(G)\le\gamma_2(G)$, and hence the theorem follows.
	\end{proof}

	\section{Dual-server domination in block graphs}
	
	By Theorem~\ref{Dserver:np_complete}, \textsc{DualServer} is NP-complete on undirected path graphs. In this section, we study the parameter for block graphs, which is a proper subclass of undirected path graphs. We show that $\gamma_2(G)=\gds(G)$ for every block graph $G$. Lan and Chang~\cite{LanChang2013} proved that $\gamma_2(G)$ can be computed in linear time for each block graph $G$. Consequently, $\gds(G)$ can be computed in linear time for each block graph $G$.
	
	Recall that a \emph{block} of a graph $G$ is a maximal connected subgraph of $G$ having no cut vertex, and that $G$ is a \emph{block graph} if every block of $G$ is a complete graph. In this section, we prove that $\gamma_{{ds}}(G)=\gamma_2(G)$ for block graphs.
	
	\begin{theorem}\label{thm:block}
		If $G$ is a block graph, then $\gamma_{{ds}}(G)=\gamma_{2}(G)$.
	\end{theorem}
	
	\begin{proof}
		Let $G=(V,E)$ be a block graph. By Observation~\ref{prop:twodom}, we have
		$\gamma_{2}(G)\le\gamma_{{ds}}(G)$, and so it suffices to show that
		$\gamma_{{ds}}(G)\le\gamma_{2}(G)$. Let $D$ be a $\gamma_{2}$-set of $G$. If
		$V-D=\emptyset$, then $D=V$ is a dual-server dominating set of $G$ and we are done.
		Hence, we may assume that $V-D\ne\emptyset$, and we recall that every vertex of $V-D$ has at least two neighbors in $D$. For every block $B$ of $G$ with $|D\cap B|\ge 2$, we fix two distinct vertices $x_{B},y_{B}\in D\cap B$. We now construct a spanning subgraph $F$ of $G$ with $V(F)=V$ as follows: we delete all the edges of $G[D]$, we delete all the edges
		of $G[V-D]$, and at each vertex $v\in V-D$ we retain exactly two of the edges joining $v$ to $D$, chosen according to the following two cases.

		\medskip
		
		\noindent \emph{Case A: Some block of $G$ containing $v$ has at least two vertices of
			$D$.}
		
		\medskip
		
		Let $B_{v}$ denote one such block. We retain the two edges $vx_{B_{v}}$ and $vy_{B_{v}}$. Note that these edges are edges of $G$ since the block $B_{v}$ is complete and contains the three vertices $v$, $x_{B_{v}}$ and $y_{B_{v}}$.
		
		\medskip
		
		\noindent\emph{Case B: Every block of $G$ containing $v$ has at most one vertex of $D$.}
		
		\medskip
		
		Since $D$ is a $2$-dominating set of $G$, we have $|N(v)\cap D|\ge 2$. Every neighbor of $v$ lies in a block containing $v$, and so, by the assumption of this case, $v$ has two neighbors in $D$ belonging to distinct blocks of $G$. We retain the two edges joining $v$ to two such neighbors.
		
		\smallskip
		It can be seen that the sets $D$ and $V-D$ are independent sets in the graph $F$, and each vertex in $V-D$ has exactly two neighbors in $D$.
		Next we construct the graph $G_{D}$ as follows: $V(G_{D})=D$, and two vertices $x$ and $y$ of $V(G_{D})$ are adjacent in $G_{D}$ if there is a vertex $v\in V-D$ adjacent to both $x$ and $y$ in $F$. Since in $F$, every vertex of $V-D$ has exactly two neighbors in $D$, each vertex of $V-D$ determines exactly one edge of $G_{D}$, and every edge of $G_{D}$ arises in this way. Note that distinct vertices of $V-D$ may, however,
		determine the same edge of $G_{D}$.
		
		\begin{claim}
			\label{cl:block1}
			The graph $G_{D}$ contains no cycle.
		\end{claim}
		
		\begin{proof}[Proof of Claim~\ref{cl:block1}]
			Suppose, to the contrary, that $d_{1}d_{2}\cdots d_{k}d_{1}$ is a cycle of $G_{D}$, where $k\ge 3$. For each $i\in[k]$, let $v_{i}\in V-D$ be a vertex whose two neighbors in $F$ are $d_{i}$ and $d_{i+1}$, where indices are taken modulo $k$. Note that the vertices $v_{1},\ldots,v_{k}$ are pairwise distinct since each vertex of $V-D$ determines exactly one edge of $G_{D}$.
			Further, $v_{i}\ne d_{j}$ for all $i,j\in[k]$, since
			$v_{i}\notin D$ while $d_{j}\in D$. In $G$, $v_{i}$ is adjacent to both $d_{i}$ and $d_{i+1}$ for each $i\in[k]$. Thus, it follows that
			$d_{1}v_{1}d_{2}v_{2}\cdots d_{k}v_{k}d_{1}$ is a cycle of $G$ of length $2k\ge 6$. Since every cycle of $G$ is contained in a single block of $G$, there is a block $B_{0}$ of $G$ with $d_{1},\ldots,d_{k},v_{1},\ldots,v_{k}\in V(B_{0})$. Let $t\in[k]$. Since $v_{t}\in V(B_{0})$ and $d_{t},d_{t+1}$ are two distinct vertices of $D\cap V(B_{0})$, by Case~A, we have $\{d_{t},d_{t+1}\}=\{x_{B_{v_{t}}},y_{B_{v_{t}}}\}\subseteq V(B_{v_{t}})$.
			Thus the blocks $B_{v_{t}}$ and $B_{0}$ have the two vertices $d_{t}$ and
			$d_{t+1}$ in common, implying that $B_{v_{t}}=B_{0}$ and
			$\{d_{t},d_{t+1}\}=\{x_{B_{0}},y_{B_{0}}\}$. Since  $t$ is arbitrary, it follows that $\{d_{i},d_{i+1}\}=\{x_{B_{0}},y_{B_{0}}\}$ for each $i\in[k]$. This is a contradiction to the fact that $d_{1},d_{2},\ldots, d_{k}$ are distinct vertices and $k\ge 3$. This proves the claim.
		\end{proof}

		By Claim~\ref{cl:block1}, the graph $G_{D}$ is a forest and thus a bipartite graph. Let $\{R,B\}$ be a bipartition of $G_{D}$, where the vertices of $D$ that are isolated in $G_{D}$ are placed in either set. It follows from the construction that each vertex in $V-D$ has a neighbor in $R$ and another neighbor in $B$ in $F$ (and thus in $G$). Therefore $D$, with associated partition $\{R,B\}$, is a dual-server dominating set of $G$, implying that $\gamma_{ds}(G)\le|D|=\gamma_{2}(G)$.
	\end{proof}

	\section{Dual-server domination in $P_4$-sparse graphs}\label{sec:p4sparse}
	
	A graph $G$ is \emph{$P_4$-sparse} if every set of five vertices of $G$ induces at most one $P_4$. A graph $G$ is $P_4$-laden if every induced subgraph $F$ of $G$ with at most six vertices satisfies: if $F$ contains more than two induced $P_4$'s, then $F$ is a split graph~\cite{Giakoumakis1996}. By Theorem~\ref{split:np_complete}, \textsc{DualServer} is NP-complete on split graphs. Since every split graph is a $P_4$-laden graph, \textsc{DualServer} is NP-complete on $P_4$-laden graphs. In this section, we study the parameter for $P_4$-sparse graphs, which is a proper subclass of $P_4$-laden graphs. We determine the dual-server domination number of every $P_4$-sparse graph. The class of $P_4$-sparse graphs contains the class of $P_4$-free graphs. In contrast with block graphs, the equality $\gamma_{ds}=\gamma_{2}$ does not hold for $P_4$-sparse graphs. For example, consider the $P_4$-sparse graph $G = 2K_{2}+2K_{2}$. We note that $\gamma_{2}(G)=3$ and $\gamma_{ds}(G)=4$.
	
	For graphs $G_1 = (V_1,E_1)$ and $G_2 = (V_2,E_2)$ with $V_1 \cap V_2 =
	\emptyset$, the \emph{disjoint union} $G_1 \cup G_2$ is the graph with vertex set $V_1 \cup V_2$ and edge set $E_1 \cup E_2$, and the \emph{join} $G_1 + G_2$ is the graph with vertex set $V_1 \cup V_2$ and edge set $E_1 \cup E_2 \cup \{xy : x \in V_1,\ y \in V_2\}$. Both operations are associative, and we write $G_1 + G_2 + \cdots + G_l$ for the join of $l$ graphs. A graph $G$ is the join of two graphs if and only if $\overline{G}$ is disconnected.
	
	A graph $G$ is a \emph{spider} if $V(G)$ admits a partition $\{L, C, Q\}$,
	where $L = \{\ell_1,\ldots ,\ell_k\}$ and $C = \{c_1,\ldots ,c_k\}$ with
	$k \ge 2$, such that
	\begin{itemize}
		\item $L$ is an independent set and $C$ is a clique of $G$;
		\item $Q$ may be empty, on the other hand if $Q$ is not empty, then
		every vertex of $Q$ is adjacent to every vertex of $C$ and to no vertex
		of $L$;
		\item either $\ell_i c_j \in E(G)$ if and only if $i = j$ (in which
		case $G$ is a \emph{thin} spider), or $\ell_i c_j \in E(G)$ if and only
		if $i \ne j$ (in which case $G$ is a \emph{thick} spider).
	\end{itemize}
	The triple $(L,C,Q)$ is called \emph{spider partition} of $G$. We refer to
	$k = |L| = |C|$ as the \emph{size} of the spider. Note that for $k = 2$ the
	thin and the thick spider of size $k$ coincide. We assume $k \ge 3$ whenever a spider is called thick. For a spider $G$ with spider partition $(L,C,Q)$, we call the vertices of $L$, $C$ and $Q$ the \emph{legs}, the \emph{body vertices} and the \emph{head vertices} of $G$, respectively.
	
	The following structural description of $P_4$-sparse graphs is due to Jamison
	and Olariu~\cite{JO92tree}.
	
	\begin{theorem}[Jamison, Olariu~\cite{JO92tree}]\label{theorem:JO}
		Let $G$ be a $P_4$-sparse graph with $|V(G)| \ge 2$. Then exactly one of the
		following holds:
		\begin{enumerate}
			
			\item $G$ is disconnected;
			\item $\overline{G}$ is disconnected;
			\item $G$ is a spider with spider partition $(L,C,Q)$ such that $Q$ is
			empty or induces a $P_4$-sparse graph.
		\end{enumerate}
	\end{theorem}

	Theorem~\ref{theorem:JO} yields a rooted tree representation of a $P_4$-sparse graph $G$, unique up to isomorphism, which we denote by $T_G$ and call the \emph{ps-tree} of $G$. Each vertex of $G$ corresponds to a leaf of $T_G$. Each internal node $u$ of $T_G$ carries a label $\cup$, $+$ or $\odot$. Let $G_u$ denote the subgraph of $G$ induced by the leaves of the maximal subtree of $T_G$ rooted at $u$, so that $G$ itself is represented by the root of $T_G$. If $u$ has label $\cup$, then $G_u$ is the disjoint union of the graphs $G_v$ taken over the children $v$ of $u$, if $u$ has label $+$, then $G_u$ is the join of the graphs $G_v$ taken over the children $v$ of $u$, and if $u$ has label $\odot$, then $G_u$ is a spider. If $u$ has label $\odot$, then the two children of $u$ correspond to the
	graphs $G_1$ and $G_2$ from which $G_u$ arises by the spider composition~\cite{JO92tree}, where $G_1$ is an edgeless graph consisting of all legs of $G_u$ but one, and $G_2$ is the subgraph of $G_u$ induced by the remaining leg, the body and the head. The ps-tree of a $P_4$-sparse graph can be constructed in linear time~\cite{JO92recog}. In particular, the spider partition of $G_u$, and hence the type, the size and the order of the head of $G_u$, can be read off from the ps-tree in linear time.
	
	Whenever $G$ is a connected graph whose complement is disconnected we write
	\begin{equation}\label{eq:decomp}
		G \;=\; G_1 + G_2 + \cdots + G_l, \qquad l \ge 2 ,
	\end{equation}
	where $G_1,\ldots ,G_l$ are the subgraphs of $G$ induced by the components of $\overline{G}$. If $G$ is $P_4$-sparse, then each $G_i$ has a connected complement, so by Theorem~\ref{theorem:JO} each $G_i$ is trivial,
	disconnected, or a spider. We refer to \eqref{eq:decomp} as the \emph{canonical decomposition} of $G$. 
	
	\begin{obs}\label{obs:gammaparts}
		If $H$ is a graph that is disconnected or a spider, then $\gamma(H) \ge 2$.
		Consequently, in the canonical decomposition \eqref{eq:decomp} of a connected $P_4$-sparse graph, we have $\gamma(G_i) = 1$ if and only if $G_i$ is trivial.
	\end{obs}

	We record two elementary facts in the following observations.  
	
	\begin{obs}\label{obs:components}
		If $H_1, H_2, \ldots ,H_r$, where $r \ge 2$, are the components of a graph $G$,
		then
		\[
		\gamma_{ds}(G) \;=\; \sum_{i=1}^{r}\gamma_{ds}(H_i).
		\]
	\end{obs}
	
	\begin{obs}\label{obs:gds2}
		Let $G$ be a graph with $|V(G)| \ge 2$. Then $\gamma_{ds}(G) \ge 2$, moreover, $\gamma_{ds}(G) = 2 $ if and only if there are two distinct vertices $u$ and $v$ of $G$ such that $V(G) \setminus \{u,v\} \subseteq N(u) \cap N(v)$.
	\end{obs}

	\begin{lemma}\label{lemma:atmostfour}
		Let $G = G_1 + \cdots + G_l$, where $l \ge 2$. If $l \ge 3$, or if $l = 2$ and
		both $G_1$ and $G_2$ are nontrivial, then $\gamma_{ds}(G) \le 4$.
	\end{lemma}
	
	\begin{proof}
		Suppose first that $l \ge 3$. If $l\ge 4$, then we choose a vertex
		$x_i \in V(G_i)$ for $i \in [4]$ and let $S = \{x_1,x_2,x_3,x_4\}$ with
		$R = \{x_1,x_2\}$ and $B = \{x_3,x_4\}$. One can check that $S$ is a dual-server dominating set of $G$ and hence $\gamma_{ds}(G) \le 4$. Suppose next that $l = 3$ and that at least one of the graphs $G_1,G_2,G_3$ is nontrivial. Relabelling if necessary, assume that $G_1$ is nontrivial. Let $u_1,u_2$ be distinct vertices of $G_1$, and let $x_2 \in V(G_2)$ and $x_3 \in V(G_3)$. Then $S = \{u_1,u_2,x_2,x_3\}$ with $R = \{u_1,x_2\}$ and $B = \{u_2,x_3\}$ is a dual-server dominating set of $G$. Therefore, $\gamma_{ds}(G) \le 4$. Suppose next that $l = 3$ and that every $G_i$ is trivial. Then $G$ is the complete graph $K_3$, and $\gamma_{ds}(G) = 2 \le 4$.

		Next, assume that $l = 2$ and both $G_1$ and $G_2$ are nontrivial. Let $u_1,u_2$ be distinct vertices of $G_1$ and let $v_1,v_2$ be distinct vertices of $G_2$. Then $S = \{u_1,u_2,v_1,v_2\}$ with $R = \{u_1,v_1\}$ and $B = \{u_2,v_2\}$ is a dual-server dominating set of $G$ and hence
		$\gamma_{ds}(G) \le 4$.
	\end{proof}

	In the next lemma, we determine the dual-server domination number of a graph which is the join of exactly two graphs, one of which is trivial. 
	\begin{lemma}\label{lemma:K1plusH}
		If $H$ is a graph with $|V(H)| \ge 2$, then
		\[
		\gamma_{{ds}}(K_1 + H) \;=\; \min\bigl\{\, \gamma(H) + 1, \; \gamma_{{ds}}(H) \,\bigr\}.
		\]
	\end{lemma}
	
	\begin{proof}
		Let $G = K_1 + H$ and $V(K_1) = \{x\}$. Clearly, $x$ is adjacent to every
		vertex of $H$. Let $D$ be a  minimum dominating set of $H$. The set
		$\{x\} \cup D$, with $R = \{x\}$ and $B = D$, is a dual-server dominating set of $G$, since every vertex of $V(H) \setminus D$ is adjacent to $x$ and has a neighbor in $D$. Hence $\gamma_{{ds}}(G) \le \gamma(H) + 1$. Next, let  $S$ be a minimum dual-server dominating set of $H$. Since $x$ is adjacent to every vertex of $S$, it follows that $S$ is a dual-server dominating set of $G$. Hence, $\gamma_{{ds}}(G) \le \gamma_{{ds}}(H)$.
		
		Let $S$ be a minimum dual-server dominating set of $G$ with associated
		partition $\{R,B\}$. Suppose first that $x \notin S$. Then $S \subseteq V(H)$ and every vertex of $V(H) \setminus S$ has both a red and a blue neighbor in $S$. Hence $S$ is a dual-server dominating set of $H$ and
		$|S| \ge \gamma_{{ds}}(H)$. Suppose next that $x \in S$. Without loss of
		generality, assume that $x \in R$. Let $S_H = S \cap V(H)$. Since $V(H) \setminus S_H= V(G)\setminus S$,  every vertex of $ V(H) \setminus S_H$ must have a blue neighbor in $S$. Since $x\in R$, we have $B \subseteq S_H$. This implies that every vertex of $ V(H) \setminus S_H$ must have a neighbor in $S_H$, implying that $S_H$ is a dominating set
		of $H$. Therefore, $|S| = 1 + |S_H| \ge 1 + \gamma(H)$. In both cases
		$|S| \ge \min\{\gamma(H)+1,\gamma_{{ds}}(H)\}$.
	\end{proof}

	We now introduce four properties and one condition that help us in identifying $P_4$-sparse graphs with a fixed dual-server domination number.
	
	\begin{defi}[Condition $N$]\label{def:N}
		A graph $H$ satisfies \emph{condition $N$} if there exist two distinct vertices $u, z \in V(H)$ such that $V(H) = N_H[u] \cup \{z\}$. 
	\end{defi}
	
	\begin{defi}[Property $D$, $T$, $T^*$, and $T^{**}$]\label{def:props}
		Let $G$ be a connected graph with canonical decomposition
		$G = G_1 + G_2 + \cdots + G_l$, where $l \ge 2$.
		\begin{itemize}
			\item $G$ has property $D$ if at least two of the graphs
			$G_1,\ldots ,G_l$ are trivial, or $\gamma_{{ds}}(G_i) = 2$ for some
			$i \in [l]$.
			\item $G$ has property $T$ if $G_i$ satisfies condition $N$ for         some $i \in [l]$.
			\item $G$ has property $T^{*}$ if $G_i$ is trivial for some $i \in [l]$ and, in addition, either $l \ge 3$ or $l=2$ and $\gamma(G_j) = 2$ for $j \in [l] \setminus \{i\}$.
			\item $G$ has property $T^{**}$ if $\gamma_{{ds}}(G_i) = 3$
			for some $i \in [l]$.
		\end{itemize}
	\end{defi}

	Note that the four properties are pairwise non-disjoint, and none of them implies any other.

	\begin{lemma}\label{lemma:two}
		Let $G$ be a connected graph with canonical decomposition
		$G = G_1 + \cdots + G_l$, where $l \ge 2$. Then $\gamma_{{ds}}(G) = 2$ if and
		only if $G$ has property $D$.
	\end{lemma}
	
	\begin{proof}
		Since $|V(G)| \ge 2$, we have $\gamma_{{ds}}(G) \ge 2$ by Observation~\ref{obs:gds2}. Suppose first that $G$ has property $D$. If $G_i$ and $G_j$ are trivial for some $i \ne j$, with $V(G_i) = \{u\}$ and
		$V(G_j) = \{v\}$, then every vertex of $V(G) \setminus \{u,v\}$ is adjacent to both $u$ and $v$ by the definition of the join, and by Observation~\ref{obs:gds2}, we have $\gamma_{{ds}}(G) = 2$. If $\gamma_{{ds}}(G_i) = 2$ for some $i \in [l]$, let $\{u,v\}$ be a minimum dual-server dominating set of $G_i$. By Observation~\ref{obs:gds2} applied to $G_i$, every vertex of $V(G_i) \setminus \{u,v\}$ is adjacent to both $u$ and $v$, and every vertex of $V(G) \setminus V(G_i)$ is adjacent to both by the definition of the join. Hence $\gamma_{{ds}}(G) = 2$.
		
		Conversely, let $\gamma_{{ds}}(G) = 2$ and let $\{u,v\}$ be a minimum dual-server dominating set of $G$. So $V(G) \setminus \{u,v\} \subseteq N(u) \cap N(v)$ by Observation~\ref{obs:gds2}. If $u \in V(G_i)$ and $v \in V(G_j)$ with $i \ne j$, then $u$ dominates $V(G_i) \setminus \{u\}$ and $v$ dominates $V(G_j) \setminus \{v\}$, so $\gamma(G_i) = \gamma(G_j) = 1$ and, by Observation~\ref{obs:gammaparts}, both $G_i$ and $G_j$ are trivial. If
		$u,v \in V(G_i)$ for some $i \in [l]$, then $V(G_i) \setminus \{u,v\} \subseteq N_{G_i}(u) \cap N_{G_i}(v)$, and Observation~\ref{obs:gds2} applied to $G_i$ gives $\gamma_{{ds}}(G_i) = 2$. In either case, $G$ has property $D$.
	\end{proof}
	
	\begin{lemma}\label{lemma:three}
		Let $G$ be a connected graph with canonical decomposition
		$G = G_1 + \cdots + G_l$, where $l \ge 2$, and suppose that $G$ does not have property $D$. Then $\gamma_{{ds}}(G) = 3$ if and only if $G$ has at least one of the properties $T$, $T^{*}$ and $T^{**}$.
	\end{lemma}
	
	\begin{proof}
		By Lemma~\ref{lemma:two} we have $\gamma_{{ds}}(G) \ge 3$. So it suffices to
		decide when $G$ has a dual-server dominating set of cardinality $3$.
		
		\medskip
		
		\noindent \emph{Sufficiency.} Suppose that $G$ has property $T$, and let $G_i$ satisfy condition $N$, say $V(G_i) = N_{G_i}[u] \cup \{z\}$ with $u \ne z$. Choose an index $j \ne i$ and a vertex $b \in V(G_j)$, and let $S = \{u,z,b\}$ with $R = \{u\}$ and $B = \{z,b\}$. If $w \in V(G_i) \setminus \{u,z\}$, then $w$ is adjacent to $u$ by condition $N$ and to $b$ by the definition of the join. If $w \in V(G) \setminus (V(G_i) \cup \{b\})$, then $w$ is adjacent to both $u$ and $z$. Hence $S$ is a dual-server dominating set of $G$ and $\gamma_{{ds}}(G) = 3$.
		
		Suppose next that $G$ has property $T^{*}$, say $G_t$ is trivial with
		$V(G_t) = \{x\}$. If $l \ge 3$, choose $b_1 \in V(G_{j_1})$ and
		$b_2 \in V(G_{j_2})$ for two distinct indices $j_1,j_2 \in [l] \setminus
		\{t\}$, and let $S = \{x,b_1,b_2\}$ with $R = \{x\}$ and $B = \{b_1,b_2\}$.
		Every vertex of $V(G)\setminus S$ is adjacent to $x$, and it belongs to at most one of $G_{j_1}$ and $G_{j_2}$ and hence is adjacent to at least one of $b_1$ and $b_2$. So $S$ is a dual-server dominating set of $G$. If $l=2$ and $\gamma(G_j) = 2$ for $j \ne t$, let $\{b_1,b_2\}$ be a minimum dominating set of $G_j$ and let $S = \{x,b_1,b_2\}$ with
		$R = \{x\}$ and $B = \{b_1,b_2\}$. Note that $V(G)\setminus S=V(G_j)\setminus \{b_1,b_2\}$. Clearly, every vertex of $V(G)\setminus S$ is adjacent to $x$ and has a neighbor in $\{b_1,b_2\}$ since $\{b_1,b_2\}$
		dominates $G_j$. So $S$ is a dual-server dominating set of $G$. In both cases $\gamma_{{ds}}(G) = 3$.
		
		Finally, suppose that $G$ has property $T^{**}$, say $\gamma_{{ds}}(G_i) = 3$, and let $S$ be a minimum dual-server dominating set of $G_i$ with associated partition $\{R,B\}$. It is easy to see that $S$ is a dual-server dominating set of $G$ and hence $\gamma_{{ds}}(G) = 3$.
		
		\medskip 
		\noindent\emph{Necessity.} Assume that $\gamma_{{ds}}(G) = 3$ and let $S$ be a minimum dual-server dominating set of $G$ with associated partition $\{R,B\}$. Interchanging the two colors if necessary, we may assume that $|R| = 1$ and $|B| = 2$, say $R = \{u\}$ and $B = \{v,w\}$. Let $a \in [l]$ be the index with $u \in V(G_a)$. We distinguish three cases.
		
		\medskip
		
		\noindent\emph{Case 1: $v,w \in V(G_a)$.} 
		
		\medskip
		
		Then $S \subseteq V(G_a)$. Every vertex of $V(G_a) \setminus S$ has a red and a blue neighbor in $S$. Since $S \subseteq V(G_a)$, $S$ with the same partition $\{R,B\}$ is a dual-server dominating set of $G_a$ and hence $\gamma_{{ds}}(G_a) \le 3$. Since $G$ does not have property $D$, we have $\gamma_{{ds}}(G_a) \ne 2$. So $\gamma_{{ds}}(G_a) = 3$ and $G$ has property $T^{**}$.
		
		\medskip
		
		\noindent\emph{Case 2: exactly one of $v$ and $w$ lies in $V(G_a)$.} 
		
		\medskip
		
		Without loss of generality, assume that $v \in V(G_a)$ and $w \in V(G_c)$ with $c \ne a$. Let $y \in V(G_a) \setminus \{u,v\}$. Then $y \in V(G) \setminus S$, so $y$ has a red neighbor in $S$. Since $u$ is the only red vertex, $y$ is adjacent to $u$. So $V(G_a) = N_{G_a}[u] \cup \{v\}$. So $G_a$ satisfies condition $N$ and $G$ has property $T$.
		
		\medskip
		
		\noindent\emph{Case 3: $v,w \notin V(G_a)$.} 
		
		\medskip
		
		Every vertex of $V(G_a) \setminus \{u\}$ belongs to $V(G) \setminus S$ and hence has a red neighbor in $S$, which must be $u$. So $u$ dominates $G_a$, that is, $\gamma(G_a) = 1$, and $G_a$ is trivial by Observation~\ref{obs:gammaparts}. If $v$ and $w$ lie in distinct graphs $G_{c_1}$ and $G_{c_2}$, then $a$, $c_1$ and $c_2$ are three distinct indices. So $l \ge 3$ and $G$ has property $T^{*}$. Suppose next that $v,w \in V(G_c)$ for some $c \ne a$. If $l \ge 3$, then $G$ has property $T^{*}$. So assume that $l = 2$, and hence $c$ is the unique index in $[l] \setminus \{a\}$. Every vertex of $V(G_c) \setminus \{v,w\}$ belongs to $V(G) \setminus S$ and hence has a blue neighbor in $S$. Since $B = \{v,w\} \subseteq V(G_c)$, $\{v,w\}$ is a dominating set of $G_c$ and $\gamma(G_c) \le 2$. Since $v$ and $w$ are distinct vertices of $G_c$, the graph $G_c$ is nontrivial, and so $\gamma(G_c) = 2$ by Observation~\ref{obs:gammaparts}. Hence $G$ has property $T^{*}$.
	\end{proof}
	
	\begin{lemma}\label{lemma:four}
		Let $G$ be a connected graph with canonical decomposition
		$G = G_1 + \cdots + G_l$, where $l \ge 2$, and suppose that $G$ has none of the properties $D$, $T$, $T^{*}$ and $T^{**}$. Then the following hold.
		\begin{enumerate}
			\item If $l \ge 3$, or if $l = 2$ and both $G_1$ and $G_2$ are
			nontrivial, then $\gamma_{{ds}}(G) = 4$.
			\item If $l = 2$ and, say, $G_1$ is trivial, then
			\[
			\gamma_{{ds}}(G) \;=\; \min\{\gamma(G_2) + 1,\; \gamma_{{ds}}(G_2)\}
			\;\ge\; 4 .
			\]
		\end{enumerate}
	\end{lemma}
	
	\begin{proof}
		By Lemmas~\ref{lemma:two} and~\ref{lemma:three}, we have $\gamma_{{ds}}(G) \ge 4$. So (a) follows from Lemma~\ref{lemma:atmostfour}. For (b), note that $G = K_1 + G_2$. Moreover, $G_2$ is nontrivial, since otherwise $G = K_2$ has property $D$. So by Lemma~\ref{lemma:K1plusH}, $\gamma_{{ds}}(G) = \min\{\gamma(G_2)+1,\gamma_{{ds}}(G_2)\}$, and this value is at least $4$.
	\end{proof}
	
	Note that the value in (b) of Lemma~\ref{lemma:four} is not bounded above by any constant. For example, for $s \ge 4$ the star $K_{1,s} = K_1 + \overline{K_s}$ has none of the four properties, and $\gamma_{{ds}}(K_{1,s}) = \min\{s+1,s\} = s$.
	
	In the following observation, we record the domination number of a spider and decide when a spider satisfies condition $N$, which are needed for the algorithm at the end of this section. 
	
	\begin{obs}\label{obs:spider}
		Let $G$ be a spider with spider partition $(L,C,Q)$ and size $k$. Then the following hold.
		\begin{enumerate}
			\item If $G$ is thin, then $\gamma(G) = k$ and $G$ satisfies condition $N$ if and only if $k = 2$.
			\item If $G$ is thick, then $\gamma(G) = 2$ and $G$ satisfies condition $N$.
		\end{enumerate}
	\end{obs}
	
	We now determine the dual-server domination number of a spider.
	
	\begin{lemma}\label{lemma:spider}
		Let $G$ be a spider with spider partition $(L,C,Q)$ and size $k$. Then the following hold.
		\begin{enumerate}
			
			\item If $G$ is thin, then $\gamma_{{ds}}(G) = k+1$ when $|Q| \le 1$, and
			$\gamma_{{ds}}(G) = k+2$ when $|Q| \ge 2$.
			\item If $G$ is thick, then $\gamma_{{ds}}(G) = 4$.
		\end{enumerate}
	\end{lemma}
	
	\begin{proof}
		
		(a) Let $G$ be thin. Each $\ell_i$ has degree $1$ in $G$, so $\ell_i$ belongs to every dual-server dominating set of $G$. Hence $L$ is contained in every dual-server dominating set of $G$, and $\gamma_{{ds}}(G) \ge k$. The set $L$ itself is not a dual-server dominating set of $G$, since $\ell_1$ is the only neighbor of $c_1$ in $L$. Hence $\gamma_{{ds}}(G) \ge k+1$.
		
		Suppose that $|Q| \le 1$. If $Q = \emptyset$, let $S = L \cup \{c_1\}$ with $R = L$ and $B = \{c_1\}$. Then $V(G) \setminus S = C \setminus \{c_1\}$, and each $c_j$ with $j \ge 2$ is adjacent to $\ell_j \in R$ and $c_1 \in B$. If $Q = \{q\}$, let $S = L \cup \{q\}$ with $R = L$ and $B = \{q\}$. Then $V(G) \setminus S = C$, and each $c_j$ is adjacent to $\ell_j \in R$ and $q \in B$. In both cases $S$ is a dual-server dominating set of $G$, so $\gamma_{{ds}}(G) \le k+1$. Hence $\gamma_{{ds}}(G) = k+1$.
		
		Suppose now that $|Q| \ge 2$. Let $S = L \cup \{c_1,c_2\}$ with $R = L \cup \{c_1\}$ and $B = \{c_2\}$. Each $c_j$ with $j \ge 3$ is adjacent to $\ell_j \in R$ and $c_2 \in B$, and each vertex of $Q$ is adjacent to $c_1 \in R$ and $c_2 \in B$. So $S$ is a dual-server dominating set of $G$, and $\gamma_{{ds}}(G) \le k+2$. Suppose, to the contrary, that $\gamma_{{ds}}(G) = k+1$, and let $S$ be a minimum dual-server dominating set of $G$. As shown above, $L \subseteq S$, so $S = L \cup \{y\}$ for some vertex $y \in C \cup Q$. Since $|Q| \ge 2$, there is a vertex $q \in Q \setminus S$. No neighbor of $q$ lies in $L$, so $y$ is the only possible neighbor of $q$ in $S$. So $q$ does not have both a red and a blue neighbor in $S$, a contradiction. Hence $\gamma_{{ds}}(G) = k+2$.
		
		\medskip
		(b) Let $G$ be thick, so $k \ge 3$. Suppose, to the contrary, that $G$ has a dual-server dominating set $S$, with associated partition $\{R,B\}$, such that $|S| \le 3$. Let $C_S = C \cap S$. Since $L$ is independent and no head vertex is adjacent to a leg, the set of neighbors of a leg $\ell_i$ in $S$ is exactly $C_S \setminus \{c_i\}$. Hence
		\begin{equation}\label{eq:thick}
			\text{for every $i \in [k]$ with $\ell_i \notin S$, the set }
			C_S \setminus \{c_i\} \text{ contains a red and a blue vertex.}
		\end{equation}
		We distinguish three cases according to $|C_S|$. If $|C_S| \le 1$, then by \eqref{eq:thick}, $\ell_i \in S$ for every $i \in [k]$. So $k \le |S| \le 3$, and hence $k = 3$ and $S = L$. Now $c_1 \notin S$, and the neighbors of $c_1$ in $S$ are $\ell_2$ and $\ell_3$; similarly for $c_2$ and $c_3$. So each of the pairs $\{\ell_2,\ell_3\}$, $\{\ell_1,\ell_3\}$ and $\{\ell_1,\ell_2\}$ contains a red and a blue vertex, which is impossible since two of $\ell_1,\ell_2,\ell_3$ receive the same colour. If $|C_S| = 2$, say $C_S = \{c_a,c_b\}$, then $C_S \setminus \{c_a\}$ consists of a single vertex, so by \eqref{eq:thick}, $\ell_a \in S$. Similarly, $\ell_b \in S$. Hence $|S| \ge 4$, a contradiction. If $|C_S| = 3$, then $S = C_S = \{c_a,c_b,c_e\}$ and no leg belongs to $S$. So by \eqref{eq:thick}, each of the pairs $\{c_b,c_e\}$, $\{c_a,c_e\}$ and $\{c_a,c_b\}$ contains a red and a blue vertex, which is again impossible. Hence $\gamma_{{ds}}(G) \ge 4$.
		
		For the upper bound, suppose first that $k \ge 4$ and let $S = \{c_1,c_2,c_3,c_4\}$ with $R = \{c_1,c_2\}$ and $B = \{c_3,c_4\}$. Each leg is nonadjacent to at most one vertex of $S$, and hence is adjacent to a red and to a blue vertex of $S$, and every vertex of $(C \cup Q) \setminus S$ is adjacent to every vertex of $S$. Suppose next that $k = 3$ and let $S = \{c_1,c_2,c_3,\ell_3\}$ with $R = \{c_1,c_2\}$ and $B = \{c_3,\ell_3\}$. Then $\ell_1$ is adjacent to $c_2 \in R$ and $c_3 \in B$, $\ell_2$ is adjacent to $c_1 \in R$ and $c_3 \in B$, and every head vertex is adjacent to $c_1 \in R$ and $c_3 \in B$. In both cases $S$ is a dual-server dominating set of $G$, so $\gamma_{{ds}}(G) \le 4$. Hence $\gamma_{{ds}}(G) = 4$.
	\end{proof}
	
	Combining Theorem~\ref{theorem:JO} with Lemmas~\ref{lemma:two}, \ref{lemma:three}, \ref{lemma:four} and~\ref{lemma:spider}, we obtain the  following characterisation.
	
	\begin{theorem}\label{theorem:p4sparse}
		Let $G$ be a connected $P_4$-sparse graph with $|V(G)| \ge 2$. Then exactly one of the following two cases occurs.
		\begin{enumerate}
			\item $\overline{G}$ is connected. Then $G$ is a spider with spider partition $(L,C,Q)$ and size $k$, and
			\[
			\gamma_{{ds}}(G) =
			\begin{cases}
				k+1 & \text{if $G$ is thin and $|Q| \le 1$,}\\[2pt]
				k+2 & \text{if $G$ is thin and $|Q| \ge 2$,}\\[2pt]
				4   & \text{if $G$ is thick.}
			\end{cases}
			\]
			\item $\overline{G}$ is disconnected. Then $G$ has a canonical
			decomposition $G = G_1 + \cdots + G_l$ with $l \ge 2$, and
			\begin{enumerate}
				\item $\gamma_{{ds}}(G) = 2$ if and only if $G$ has property $D$;
				
				\item $\gamma_{{ds}}(G) = 3$ if and only if $G$ does not have property $D$ and has at least one of the properties $T$, $T^{*}$ and $T^{**}$;
				
				\item if $G$ has none of the properties $D$, $T$, $T^{*}$ and $T^{**}$, and if in addition $l \ge 3$ or both $G_1$ and $G_2$ are nontrivial, then $\gamma_{{ds}}(G) = 4$;
				
				\item if $G$ has none of the properties $D$, $T$, $T^{*}$ and $T^{**}$, and if $l = 2$ with $G_1$ trivial, then $\gamma_{{ds}}(G) = \min\{\gamma(G_2)+1,\ \gamma_{{ds}}(G_2)\} \ge 4$.
			\end{enumerate}
		\end{enumerate}
	\end{theorem}
	
	\begin{proof}
		Since $G$ is connected and $|V(G)| \ge 2$, by Theorem~\ref{theorem:JO} either $\overline{G}$ is connected and $G$ is a spider, or $\overline{G}$ is disconnected and $G$ has a canonical decomposition as in \eqref{eq:decomp}. Part (a) of the theorem follows from Lemma~\ref{lemma:spider}. Part (b) of the theorem follows from Lemmas~\ref{lemma:two}, \ref{lemma:three} and~\ref{lemma:four}. Note that the four subcases of (b) are mutually exclusive and exhaustive. 
	\end{proof}

	We next show that every positive integer is the dual-server domination number of some connected $P_4$-sparse graph. 
	
	\begin{theorem}\label{theorem:p4realiz}
		For every positive integer $k$,  there is a connected $P_4$-sparse graph $G$ with $\gamma_{{ds}}(G) = k$. 
	\end{theorem}
	
	\begin{proof}
		For $k = 1$ take $G = K_1$ and for $k = 2$ take $G = K_2$. For $k \ge 3$, let $G$ be the thin spider of size $k-1 \ge 2$ with empty head, and let $(L,C,\emptyset)$ be its spider partition. Then $G$ is a connected $P_4$-sparse graph, and $\gamma_{{ds}}(G) = (k-1)+1 = k$ by Lemma~\ref{lemma:spider}(a).
	\end{proof}

	We now design a linear time algorithm to compute $\gamma_{{ds}}(G)$ for a $P_4$-sparse graph $G$. By Observation~\ref{obs:components}, it suffices to consider connected $P_4$-sparse graphs. Theorem~\ref{theorem:p4sparse} determines $\gamma_{{ds}}$ at an internal node of the ps-tree from a constant amount of information attached to each of its children. Accordingly, for every node $u$ of $T_G$ we compute the quadruple
	\[
	\bigl(\, n(u),\; \gamma(u),\; \gamma_{{ds}}(u),\; \nu(u) \,\bigr),
	\]
	where $n(u) = |V(G_u)|$, $\gamma(u) = \gamma(G_u)$, $\gamma_{{ds}}(u) = \gamma_{{ds}}(G_u)$, and $\nu(u)$ is {\sc true} if $G_u$ satisfies the condition $N$ and {\sc false} otherwise. For a node $u$ of $T_G$, we denote by $C(T_G,u)$ the set of children of $u$ in $T_G$. The quadruples are computed bottom-up by the following rules, where $C(T_G,u) = \{c_1,\ldots ,c_l\}$.
	
	\begin{itemize}
		\item \emph{$u$ is a leaf.} Then $(n(u),\gamma(u),\gamma_{{ds}}(u),\nu(u)) = (1,1,1,\textsc{false})$, since a trivial graph does not satisfy condition $N$.
		\item \emph{$u$ has label $\cup$.} Then $n(u) = \sum_{j} n(c_j)$, $\gamma(u) = \sum_{j} \gamma(c_j)$, and $\gamma_{{ds}}(u) = \sum_{j} \gamma_{{ds}}(c_j)$ by Observation~\ref{obs:components}. Moreover, $\nu(u)$ is {\sc true} if and only if $l = 2$ and, for some $j \in [2]$, $n(c_j) = 1$ and $\gamma(c_{j'}) = 1$, where $j' \in [2] \setminus \{j\}$. Indeed, if $V(G_u) = N_{G_u}[x] \cup \{z\}$ with $x \ne z$, then $N_{G_u}[x]$ lies in the component of $G_u$ containing $x$, so $G_u$ has exactly two components, one of which is the trivial graph on $z$ and the other of which is dominated by $x$.
		\item \emph{$u$ has label $\odot$.} Then $G_u$ is a spider. Let $c_1$ and $c_2$ be the children of $u$ corresponding to the edgeless graph on all legs but one and to the graph induced by the remaining leg, the body and the head, respectively. Then the size of $G_u$ is $k = n(c_1) + 1$, the order of its head $Q$ is $|Q| = n(c_2) - k - 1$, and $n(u) = 2k + |Q|$. The values $\gamma(u)$, $\gamma_{{ds}}(u)$ and $\nu(u)$ are given by Lemma~\ref{lemma:spider} and Observation~\ref{obs:spider}, since each of them depends only on the type of $G_u$, on $k$, and on whether $|Q| \le 1$.
		\item \emph{$u$ has label $+$.} Then $G_u = G_{c_1} + \cdots + G_{c_l}$ is the canonical decomposition of $G_u$, and $n(u) = \sum_{j} n(c_j)$. By Observation~\ref{obs:gammaparts}, $G_{c_j}$ is trivial if and only if $n(c_j) = 1$, and $\gamma(u) = 1$ if $n(c_j) = 1$ for some $j \in [l]$, while $\gamma(u) = 2$ otherwise. Moreover, $\nu(u)$ is {\sc true} if and only if $\gamma(u) = 1$ or $\nu(c_j)$ is {\sc true} for some $j \in [l]$. Indeed, a dominating vertex $x$ of $G_u$ satisfies $V(G_u) = N_{G_u}[x] \cup \{z\}$ for any $z \ne x$, and if $V(G_{c_j}) = N_{G_{c_j}}[x] \cup \{z\}$, then $V(G_u) = N_{G_u}[x] \cup \{z\}$, since $x$ is adjacent to every vertex of $V(G_u) \setminus V(G_{c_j})$. Conversely, if $V(G_u) = N_{G_u}[x] \cup \{z\}$ with $x \ne z$ and $x \in V(G_{c_j})$, then either $z \notin V(G_{c_j})$, in which case $x$ dominates $G_{c_j}$, so $G_{c_j}$ is trivial and $\gamma(u) = 1$, or $z \in V(G_{c_j})$, in which case $G_{c_j}$ satisfies condition $N$. Finally, $\gamma_{{ds}}(u)$ is computed by Theorem~\ref{theorem:p4sparse}(b), where the properties $D$, $T$, $T^{*}$ and $T^{**}$ are decided from the quadruples of the children as follows. The graph $G_u$ has property $D$ if and only if $n(c_j) = 1$ for at least two indices $j \in [l]$ or $\gamma_{{ds}}(c_j) = 2$ for some $j \in [l]$. It has property $T$ if and only if $\nu(c_j)$ is {\sc true} for some $j \in [l]$. It has property $T^{*}$ if and only if $n(c_j) = 1$ for some $j \in [l]$ and, in addition, either $l \ge 3$ or $l = 2$ and $\gamma(c_{j'}) = 2$ for $j' \in [l] \setminus \{j\}$. Finally, it has property $T^{**}$ if and only if $\gamma_{{ds}}(c_j) = 3$ for some $j \in [l]$.
	\end{itemize}
	
	If we apply breadth first search at the root $u_r$ of $T_G$, then we get an ordering, say $\phi$, of the internal nodes of $T_G$. We use the ordering $\sigma = (u_1,u_2,\ldots ,u_r)$, where $\sigma = \phi^{-1}$, and call $\sigma$ an IRBFS ordering of $T_G$. In $\sigma$ every internal node is preceded by all of its internal children. We are now in a position to present Algorithm~\ref{p4sparse-algo} for computing $\gamma_{{ds}}(G)$ in a given connected $P_4$-sparse graph $G$.
	
	\begin{algorithm}[H]
		\SetAlgoVlined
		\KwIn{A connected $P_4$-sparse graph $G$\;}
		\KwOut{$\gamma_{{ds}}(G)$\;}
		Construct the ps-tree $T_G$ of $G$\;
		Compute an IRBFS ordering $\sigma=(u_1, u_2, \ldots, u_r)$ of the ps-tree $T_G$ of $G$\;
		\
		Initialize $(n(v),\gamma(v),\gamma_{{ds}}(v),\nu(v)) = (1,1,1,\textsc{false})$ for each leaf $v$ of $T_G$\;
		
		\For{$(i=1$ to $r)$}
		{
			Let $C(T_G, u_i)=\{c_1, c_2, \ldots, c_l\}$\;
			$n(u_i) = \displaystyle\sum_{j=1}^{l} n(c_j)$\;
			\uIf {$(u_i$ is a node with label $\cup)$}
			{
				$\gamma(u_i)=\displaystyle \sum_{j=1}^{l}\gamma(c_j)$ and $\gamma_{{ds}}(u_i)=\displaystyle \sum_{j=1}^{l}\gamma_{{ds}}(c_j)$\;
				$\nu(u_i) = \textsc{true}$ if $l = 2$, $n(c_j) = 1$ and $\gamma(c_{j'}) = 1$ for $j,j' \in [2]$ with $j\neq j'$, and $\nu(u_i) = \textsc{false}$ otherwise\;
			}
			\uElseIf {$(u_i$ is a node with label $\odot)$}
			{
				Let $k = n(c_1)+1$ and $|Q| = n(c_2) - k - 1$, where $G_{c_1}$ is edgeless\;
				Compute $\gamma(u_i)$, $\gamma_{{ds}}(u_i)$ and $\nu(u_i)$ by Lemma~\ref{lemma:spider} and Observation~\ref{obs:spider} from the type of $G_{u_i}$, $k$ and $|Q|$\;
			}
			\Else{
				$\gamma(u_i) = 1$ if $n(c_j) = 1$ for some $j \in [l]$, and $\gamma(u_i) = 2$ otherwise\;
				$\nu(u_i) = \textsc{true}$ if $\gamma(u_i) = 1$ or $\nu(c_j) = \textsc{true}$ for some $j \in [l]$, and $\nu(u_i) = \textsc{false}$ otherwise\;
				\uIf{$(G_{u_i}$ has property $D)$}{$\gamma_{{ds}}(u_i)=2$\;}
				\ElseIf{$(G_{u_i}$ has property $T$, $T^{*}$ or $T^{**})$}{$\gamma_{{ds}}(u_i)=3$\;}
				\ElseIf{$(l \ge 3$, or $l = 2$ and $n(c_1),n(c_2) \ge 2)$}{$\gamma_{{ds}}(u_i)=4$\;}
				\Else{$\gamma_{{ds}}(u_i)=\min\{\gamma(c_{j'})+1,\ \gamma_{{ds}}(c_{j'})\}$, where $n(c_j) = 1$ and $j,j' \in [2]$ with $j\neq j'$\;}
		}}
		
		\Return $\gamma_{{ds}}(u_r)$\;
		\caption{\label{p4sparse-algo}\textsc{DS-P4Sparse($G$)}}
	\end{algorithm}
	
	\medskip
	
	The algorithm \textsc{DS-P4Sparse($G$)} correctly computes $\gamma_{{ds}}(G)$ using Observations~\ref{obs:gammaparts}-\ref{obs:spider}, Lemma~\ref{lemma:spider} and Theorem~\ref{theorem:p4sparse}. We now show that the algorithm \textsc{DS-P4Sparse($G$)} can be executed in linear time. Let $G$ be a connected $P_4$-sparse graph with $|V(G)| = n$ and $|E(G)| = m$. The construction of the ps-tree $T_G$ of $G$ takes $O(n+m)$ time~\cite{JO92recog}, and since every internal node of $T_G$ has at least two children, $T_G$ has $O(n)$ nodes and $O(n)$ edges. Hence the IRBFS ordering $\sigma$ and the initialization of the quadruples at the leaves of $T_G$ take $O(n)$ time. At an internal node $u_i$, the entries $n(u_i)$, $\gamma(u_i)$, $\gamma_{{ds}}(u_i)$ and $\nu(u_i)$ when $u_i$ has label $\cup$, are sums or tests over the children of $u_i$. When $u_i$ has label $\odot$, its type is available from the ps-tree, and $k$ and $|Q|$ are obtained in $O(1)$ time from $n(c_1)$ and $n(c_2)$. When $u_i$ has label $+$, the values $\gamma(u_i)$ and $\nu(u_i)$, each of the properties $D$, $T$, $T^{*}$ and $T^{**}$, and whether $l \ge 3$ or both $G_{c_1}$ and $G_{c_2}$ are nontrivial, are decided by inspecting the quadruples of the children of $u_i$. In the remaining case $l = 2$ and $n(c_j) = 1$ for some $j \in [2]$, and $\gamma_{{ds}}(u_i) = \min\{\gamma(c_{j'})+1,\gamma_{{ds}}(c_{j'})\}$ is obtained in $O(1)$ time from the quadruple of the other child $c_{j'}$. So processing $u_i$ takes $O(|C(T_G,u_i)|)$ time in every case. Summing over all internal nodes of $T_G$, the ``for'' loop takes $O(n)$ time. Hence, it takes $O(n+m)$ time for the computation of $\gamma_{{ds}}(G)$ by the algorithm \textsc{DS-P4Sparse($G$)}, and the following theorem holds.
	
	\begin{theorem}\label{theorem:p4alg}
		The dual-server domination number of a $P_4$-sparse graph $G$ of order $n$ and size $m$ can be computed in $O(n+m)$ time.
	\end{theorem}

	\section{Dual-server domination in series-parallel graphs}\label{sec:sp}
	A \emph{two-terminal series-parallel graph} is a graph with two distinguished vertices, $s$ and $t$, called its terminals, constructed recursively from a single edge by series and parallel compositions. In a series composition, one terminal of $G_1$ is merged with one terminal of $G_2$. In a parallel composition, both terminals of $G_1$ are merged with the corresponding terminals of $G_2$. By Theorem~\ref{Dserver:np_complete}, \textsc{DualServer} is NP-complete on planar graphs. In this section, we study the parameter for series-parallel graphs, which is a proper subclass of planar graphs. Let $G$ be a series-parallel graph with terminals $x$ and $y$. We denote by $\mathcal{F}(G)$ the set of all triples $(S,R,B)$ such that $S\subseteq V(G)$ and $\{R,B\}$ is an associated partition of $S$, that is, $R\cap B=\emptyset$ and $R\cup B=S$. However, we allow $R$ or $B$ to be an empty set. Clearly, $S$ is a dual-server dominating set of $G$ with associated partition $\{R,B\}$ if every vertex of $V(G)-S$ has a neighbor in $R$ and a neighbor in $B$. For $i,j\in\{0,1,2,3,4,5\}$, we define the sets $\mathcal{S}(G,x_i,y_j)$ as follows.
	\begin{itemize}
		\item For $i,j\in\{0,1,2\}$,
		$\mathcal{S}(G,x_i,y_j)=\{(S,R,B)\in\mathcal{F}(G): S$ is a
		dual-server dominating set of $G$ with associated partition $\{R,B\}$; $x\in R$ if $i=1$, $x\in B$ if $i=2$, $x\notin S$ if $i=0$; and
		$y\in R$ if $j=1$, $y\in B$ if $j=2$, $y\notin S$ if $j=0\}$.
		\item For $i\in\{3,4,5\}$ and $j\in\{0,1,2\}$,
		$\mathcal{S}(G,x_i,y_j)=\{(S,R,B)\in\mathcal{F}(G): S$ is a dual-server dominating set of $G-x$ with associated partition $\{R,B\}$; $x\notin S$;
		$N_G(x)\cap R\neq\emptyset$ and $N_G(x)\cap B=\emptyset$ if $i=3$,
		$N_G(x)\cap B\neq\emptyset$ and $N_G(x)\cap R=\emptyset$ if $i=4$,
		$N_G(x)\cap S=\emptyset$ if $i=5$; and
		$y\in R$ if $j=1$, $y\in B$ if $j=2$, $y\notin S$ if $j=0\}$.
		\item For $i\in\{0,1,2\}$ and $j\in\{3,4,5\}$,
		$\mathcal{S}(G,x_i,y_j)=\{(S,R,B)\in\mathcal{F}(G): S$ is a
		dual-server dominating set of $G-y$ with associated partition
		$\{R,B\}$; $y\notin S$; $N_G(y)\cap R\neq\emptyset$ and $N_G(y)\cap B=\emptyset$ if $j=3$, $N_G(y)\cap B\neq\emptyset$ and $N_G(y)\cap R=\emptyset$ if $j=4$, $N_G(y)\cap S=\emptyset$ if $j=5$; and
		$x\in R$ if $i=1$, $x\in B$ if $i=2$, $x\notin S$ if $i=0\}$.
		\item For $i,j\in\{3,4,5\}$,
		$\mathcal{S}(G,x_i,y_j)=\{(S,R,B)\in\mathcal{F}(G): S$ is a
		dual-server dominating set of $G-\{x,y\}$ with associated partition
		$\{R,B\}$; $x,y\notin S$;  $N_G(x)\cap R\neq\emptyset$ and $N_G(x)\cap B=\emptyset$ if $i=3$, $N_G(x)\cap B\neq\emptyset$ and $N_G(x)\cap R=\emptyset$ if $i=4$, $N_G(x)\cap S=\emptyset$ if $i=5$; $N_G(y)\cap R\neq\emptyset$ and $N_G(y)\cap B=\emptyset$ if $j=3$, $N_G(y)\cap B\neq\emptyset$ and $N_G(y)\cap R=\emptyset$ if $j=4$, $N_G(y)\cap S=\emptyset$ if $j=5\}$.
	\end{itemize}

	For $i,j\in\{0,1,2,3,4,5\}$, let $\lambda(G,x_i,y_j)=\min\{|S|:(S,R,B)\in\mathcal{S}(G,x_i,y_j)\}$, where $\lambda(G,x_i,y_j)=\infty$ if $\mathcal{S}(G,x_i,y_j)=\emptyset$. The proofs of Lemma~\ref{lem:sp-root} and Lemma~\ref{lem:sp-leaf} are easy and hence are omitted.
	
	\begin{lemma}\label{lem:sp-root}
		If $G$ is a series-parallel graph with terminals $x$ and $y$, then
		$\gamma_{{ds}}(G)=\min\{\lambda(G,x_i,y_j): i,j\in\{0,1,2\}\}$.
	\end{lemma}
	
	\begin{lemma}\label{lem:sp-leaf}
		Let $G$ be a series-parallel graph with two vertices $x,y$ and an edge $xy$. Then, for $i,j\in\{0,1,2,3,4,5\}$, the following hold.
		\begin{enumerate}
			\item $\lambda(G,x_i,y_j)=2$, where $i,j\in\{1,2\}$.
			\item $\lambda(G,x_1,y_3)=\lambda(G,x_3,y_1)=
			\lambda(G,x_2,y_4)=\lambda(G,x_4,y_2)=1$.
			\item $\lambda(G,x_5,y_5)=0$.
			\item $\lambda(G,x_i,y_j)=\infty$ otherwise.
		\end{enumerate}
	\end{lemma}
	
	Let $G$ be a series-parallel graph obtained from the series or parallel
	connection of two series-parallel graphs $G_1$ and $G_2$. In the next two lemmas, we show how to obtain the value of the parameter corresponding to $G$ from those of $G_1$ and $G_2$. In both proofs, for $(S,R,B)\in\mathcal{F}(G)$ and $p\in\{1,2\}$, we write $(S_p,R_p,B_p)=(S\cap V(G_p),R\cap V(G_p),B\cap V(G_p))$ for the restriction of $(S,R,B)$ to $G_p$. Clearly, $(S_p,R_p,B_p)\in\mathcal{F}(G_p)$. Since $V(G_1)\cap V(G_2)$ consists only of terminals, every vertex of $V(G_p)-S_p$ that is not a terminal of $G_p$ has all its neighbors in $G_p$, hence it has a neighbor in $R_p$ and a neighbor in $B_p$ whenever it has a neighbor in $R$ and a neighbor in $B$. We use this fact without further mention in the rest of this section.
	
	\begin{lemma}\label{lem:sp-series}
		Suppose that a series-parallel graph $G$ is obtained from the series connection of two series-parallel graphs $G_1$ and $G_2$. Let $x,y$ be the terminals of $G_1$ and $s,t$ be the terminals of $G_2$. Then for $i,j\in\{0,1,2,3,4,5\}$,
		$$
		\lambda(G,x_i,t_j)=\min\left\{
		\begin{array}{ll}
			\lambda(G_1,x_i,y_1)+\lambda(G_2,s_1,t_j)-1, & \\[2pt]
			\lambda(G_1,x_i,y_2)+\lambda(G_2,s_2,t_j)-1, & \\[2pt]
			\lambda(G_1,x_i,y_0)+\lambda(G_2,s_k,t_j), & k\in\{0,3,4,5\},\\[2pt]
			\lambda(G_1,x_i,y_k)+\lambda(G_2,s_0,t_j), & k\in\{3,4,5\},\\[2pt]
			\lambda(G_1,x_i,y_3)+\lambda(G_2,s_4,t_j), & \\[2pt]
			\lambda(G_1,x_i,y_4)+\lambda(G_2,s_3,t_j) &
		\end{array}\right\}.
		$$
	\end{lemma}
	
	\begin{proof}
		Let $(S,R,B)\in\mathcal{S}(G,x_i,t_j)$. Since $y$ and $s$ are the same vertex in $G$, we have $y\in R_1$ if and only if $s\in R_2$, and $y\in B_1$ if and only if $s\in B_2$. If $y\in R$, then $(S_1,R_1,B_1)\in\mathcal{S}(G_1,x_i,y_1)$, $(S_2,R_2,B_2)\in\mathcal{S}(G_2,s_1,t_j)$, and $|S|=|S_1|+|S_2|-1$, as $y$ is counted in both restrictions. Analogously, if $y\in B$, then $(S_1,R_1,B_1)\in\mathcal{S}(G_1,x_i,y_2)$, $(S_2,R_2,B_2)\in\mathcal{S}(G_2,s_2,t_j)$, and $|S|=|S_1|+|S_2|-1$. Now assume that $y\notin S$. Then $y$ has a neighbor $u\in R$ and a neighbor $v\in B$, and $|S|=|S_1|+|S_2|$. Since $N_G(y)=N_{G_1}(y)\cup N_{G_2}(s)$, one of the following holds: $u,v\in V(G_1)$, in which case $(S_1,R_1,B_1)\in\mathcal{S}(G_1,x_i,y_0)$ and $(S_2,R_2,B_2)\in\mathcal{S}(G_2,s_k,t_j)$ for some $k\in\{0,3,4,5\}$; $u,v\in V(G_2)$, in which case $(S_1,R_1,B_1)\in\mathcal{S}(G_1,x_i,y_k)$ for some $k\in\{0,3,4,5\}$ and $(S_2,R_2,B_2)\in\mathcal{S}(G_2,s_0,t_j)$; $u\in V(G_1)$, $v\in V(G_2)$, $N_{G_1}(y)\cap B_1=\emptyset$ and $N_{G_2}(s)\cap R_2=\emptyset$, in which case
		$(S_1,R_1,B_1)\in\mathcal{S}(G_1,x_i,y_3)$ and $(S_2,R_2,B_2)\in\mathcal{S}(G_2,s_4,t_j)$; or, symmetrically, $(S_1,R_1,B_1)\in\mathcal{S}(G_1,x_i,y_4)$ and $(S_2,R_2,B_2)\in\mathcal{S}(G_2,s_3,t_j)$. Thus, the lemma is proved.
	\end{proof}
	
	\begin{lemma}\label{lem:sp-parallel}
		Suppose that a series-parallel graph $G$ is obtained from the parallel
		connection of two series-parallel graphs $G_1$ with terminals $x,y$ and
		$G_2$ with terminals $s,t$. Then the following hold.
		\begin{enumerate}
			\item For $i,j\in\{1,2,5\}$,
			\[
			\lambda(G,x_i,y_j)=\lambda(G_1,x_i,y_j)+\lambda(G_2,s_i,t_j)-\delta_i-\delta_j,
			\]
			where $\delta_i=1$ if $i\in\{1,2\}$ and $\delta_i=0$ if $i=5$, and analogously, $\delta_j=1$ if $j\in\{1,2\}$ and $\delta_j=0$ if $j=5$. 
			
			\item For $i\in\{1,2,5\}$ and $j\in\{3,4\}$,
			\[
			\lambda(G,x_i,y_j)=\min\left\{
			\begin{array}{l}
				\lambda(G_1,x_i,y_j)+\lambda(G_2,s_i,t_j)\\[2pt]
				\lambda(G_1,x_i,y_j)+\lambda(G_2,s_i,t_5)\\[2pt]
				\lambda(G_1,x_i,y_5)+\lambda(G_2,s_i,t_j)
			\end{array}\right\}-\delta_i,
			\]
			where $\delta_i=1$ if $i\in\{1,2\}$ and $\delta_i=0$ if $i=5$. 
			
			Symmetrically, for  $i\in\{3,4\}$ and $j\in\{1,2,5\}$,
			\[
			\lambda(G,x_i,y_j)=\min\left\{
			\begin{array}{l}
				\lambda(G_1,x_i,y_j)+\lambda(G_2,s_i,t_j)\\[2pt]
				\lambda(G_1,x_5,y_j)+\lambda(G_2,s_i,t_j)\\[2pt]
				\lambda(G_1,x_i,y_j)+\lambda(G_2,s_5,t_j)
			\end{array}\right\}-\delta_j,
			\]
			where $\delta_j=1$ if $j\in\{1,2\}$ and $\delta_j=0$ if $j=5$.

			\item For $i\in\{1,2,5\}$,
			\[
			\lambda(G,x_i,y_0)=\min\left\{
			\begin{array}{ll}
				\lambda(G_1,x_i,y_0)+\lambda(G_2,s_i,t_k), & k\in\{0,3,4,5\},\\[2pt]
				\lambda(G_1,x_i,y_k)+\lambda(G_2,s_i,t_0), & k\in\{3,4,5\},\\[2pt]
				\lambda(G_1,x_i,y_3)+\lambda(G_2,s_i,t_4), & \\[2pt]
				\lambda(G_1,x_i,y_4)+\lambda(G_2,s_i,t_3) &
			\end{array}\right\}-\delta_i,
			\]
			where $\delta_i=1$ if $i\in\{1,2\}$ and $\delta_i=0$ if $i=5$.
			
			Symmetrically, for $j\in\{1,2,5\}$, 
			\[
			\lambda(G,x_0,y_j)=\min\left\{
			\begin{array}{ll}
				\lambda(G_1,x_0,y_j)+\lambda(G_2,s_k,t_j), & k\in\{0,3,4,5\},\\[2pt]
				\lambda(G_1,x_k,y_j)+\lambda(G_2,s_0,t_j), & k\in\{3,4,5\},\\[2pt]
				\lambda(G_1,x_3,y_j)+\lambda(G_2,s_4,t_j), & \\[2pt]
				\lambda(G_1,x_4,y_j)+\lambda(G_2,s_3,t_j) &
			\end{array}\right\}-\delta_j,
			\]
			where $\delta_j=1$ if $j\in\{1,2\}$ and $\delta_j=0$ if $j=5$.

			\item For $i,j\in\{0,3,4\}$,
			\[
			\lambda(G,x_i,y_j)=\min\bigl\{\lambda(G_1,x_{i_1},y_{j_1})+\lambda(G_2,s_{i_2},t_{j_2})\bigr\},
			\]
			where the minimum is taken over all $(i_1,i_2)$ and $(j_1,j_2)$ such that
			$(i_1,i_2)\in\{(i,i),(i,5),(5,i)\}$ if $i\in\{3,4\}$,
			$(i_1,i_2)\in\{(0,0),(0,3),(0,4),(0,5),(3,0),(4,0),(5,0),(3,4),(4,3)\}$ if
			$i=0$, and $(j_1,j_2)$ satisfies the same condition with respect to $j$.
			
		\end{enumerate}
	\end{lemma}
	
	\begin{proof}
		Let $(S,R,B)\in\mathcal{S}(G,x_i,y_j)$. Since $x$ and $s$ are the same vertex of $G$, and $y$ and $t$ are the same vertex of $G$, we have
		$|S|=|S_1|+|S_2|-|S\cap\{x,y\}|$. Moreover, $N_G(x)=N_{G_1}(x)\cup N_{G_2}(s)$ and $N_G(y)=N_{G_1}(y)\cup N_{G_2}(t)$.
		
		\medskip 
		
		(a) Here $|S\cap\{x,y\}|=\delta_i+\delta_j$. If $i=1$, then $x\in R_1$ and $s\in R_2$.  If $i=2$, then $x\in B_1$ and $s\in B_2$. If $i=5$, then
		$x\notin S$ and $N_G(x)\cap S=\emptyset$, hence $N_{G_1}(x)\cap S_1=\emptyset$ and $N_{G_2}(s)\cap S_2=\emptyset$. The same holds for $y$ and $j$. Therefore $(S_1,R_1,B_1)\in\mathcal{S}(G_1,x_i,y_j)$ and
		$(S_2,R_2,B_2)\in\mathcal{S}(G_2,s_i,t_j)$. Thus (a) holds.
		
		\medskip
		(b) We only prove the case when $i\in\{1,2,5\}$ and $j\in\{3,4\}$ since the case $i\in\{3,4\}$ and $j\in\{1,2,5\}$ follows by analogous argument.
		Let $j=3$. Here $y\notin S$, so $|S\cap\{x,y\}|=\delta_i$. As in (a), the restrictions satisfy the condition on $x$ for the index $i$. Since
		$N_G(y)\cap R\neq\emptyset$ and $N_G(y)\cap B=\emptyset$, we have
		$N_{G_1}(y)\cap B_1=\emptyset$ and $N_{G_2}(t)\cap B_2=\emptyset$, while at
		least one of $N_{G_1}(y)\cap R_1$ and $N_{G_2}(t)\cap R_2$ is nonempty. Hence $(S_1,R_1,B_1)\in\mathcal{S}(G_1,x_i,y_{j_1})$ and
		$(S_2,R_2,B_2)\in\mathcal{S}(G_2,s_i,t_{j_2})$ with
		$(j_1,j_2)\in\{(3,3),(3,5),(5,3)\}$. The case $j=4$ is analogous with $B$ in place of $R$. Thus (b) holds.
		
		\medskip
		(c)  We only prove the case when $i\in\{1,2,5\}$ since the case $j\in\{1,2,5\}$ follows by analogous argument. Here $y\notin S$, so $|S\cap\{x,y\}|=\delta_i$, and $y$ has a neighbor $u\in R$ and a neighbor $v\in B$. Exactly as in the proof of Lemma~\ref{lem:sp-series}, according to whether $u,v$ lie in $G_1$, in $G_2$, or one in each, we obtain $(S_1,R_1,B_1)\in\mathcal{S}(G_1,x_i,y_{j_1})$ and $(S_2,R_2,B_2)\in\mathcal{S}(G_2,s_i,t_{j_2})$ with $(j_1,j_2)$ one of $(0,k)$
		for $k\in\{0,3,4,5\}$, $(k,0)$ for $k\in\{3,4,5\}$, $(3,4)$, or $(4,3)$. Thus (c) holds.
		
		\medskip
		(d) Here $x,y\notin S$, so $|S|=|S_1|+|S_2|$. We apply the arguments of (b) and (c) to $x$ and to $y$ independently. The argument of (b) gives the admissible pairs $(i_1,i_2)$ when $i\in\{3,4\}$ and $(j_1,j_2)$ when $j\in\{3,4\}$. The argument of (c) gives the admissible pairs  $(i_1,i_2)$ and $(j_1,j_2)$ when $i=0$ or $j=0$. Thus (d) holds.
	\end{proof}
	
	We now present a dynamic programming algorithm to compute $\gamma_{{ds}}(G)$
	for a given series-parallel graph $G$ with terminals $x$ and $y$. At first we construct a parse tree $T$ of $G$. Our algorithm traverses $T$ in a bottom-up manner. Let $\{v_1,v_2,\ldots,v_r\}$ be the vertex set of $T$ and $v_r$ be the root of $T$. We use an ordering $\sigma=(v_1,v_2,\ldots,v_r)$ of the vertices of $T$ for the traversal, where $\sigma^{-1}=(v_r,v_{r-1},\ldots,v_1)$ is an ordering obtained by performing a breadth-first search starting at $v_r$. We call such an ordering $\sigma$ a Reverse-BFS ordering of $T$. Recall that each leaf of $T$ represents an edge of $G$. If $v_l$ is a leaf of $T$ and $pq$ is the edge corresponding to $v_l$, then the values
	$\lambda(H,p_i,q_j)$, for $i,j\in\{0,1,2,3,4,5\}$, are determined by
	Lemma~\ref{lem:sp-leaf}, where $H$ denotes the series-parallel graph
	consisting of the two vertices $p,q$ and the edge $pq$. When an internal
	vertex $v_k$ of $T$ is considered, the algorithm computes the values
	$\lambda(H,p_i,q_j)$, for $i,j\in\{0,1,2,3,4,5\}$, using
	Lemma~\ref{lem:sp-series} or Lemma~\ref{lem:sp-parallel}, where $H$ is the series-parallel graph with terminals $p$ and $q$  corresponding to $v_k$.  Note that the labels of the two children of $v_k$ together with the label of $v_k$ indicate whether $H$ is obtained by a series connection or by a parallel connection. If $H$ is obtained by a parallel connection, then both children of $v_k$ are labelled $(p,q)$, and by a series connection otherwise. Accordingly, the algorithm applies Lemma~\ref{lem:sp-parallel} if $H$ is obtained by a parallel connection and applies Lemma~\ref{lem:sp-series} if $H$ is obtained by a series connection. Finally, the algorithm computes the values $\lambda(G,x_i,y_j)$, for $i,j\in\{0,1,2,3,4,5\}$, at the root of $T$ and determines $\gamma_{{ds}}(G)$ using Lemma~\ref{lem:sp-root}.
	
	\begin{algorithm}[H]
		\DontPrintSemicolon
		\caption{\textsc{DS-SP}$(G)$}\label{alg:ds-sp}
		\KwIn{A series-parallel graph $G$ with terminals $x$ and $y$.}
		\KwOut{$\gamma_{{ds}}(G)$.}
		Construct a parse tree $T$ of $G$\;
		Compute a Reverse-BFS ordering $\sigma=(v_1,v_2,\ldots,v_r)$ of $T$\;
		\For{$s=1$ \KwTo $r$}{
			Let $(p,q)$ be the label of $v_s$ and let $H$ be the series-parallel
			graph corresponding to $v_s$\;
			\uIf{$v_s$ is a leaf of $T$}{
				For $i,j\in\{0,1,2,3,4,5\}$, determine $\lambda(H,p_i,q_j)$ using
				Lemma~\ref{lem:sp-leaf}\;
			}
			\uElseIf{both children of $v_s$ have label $(p,q)$}{
				For $i,j\in\{0,1,2,3,4,5\}$, determine $\lambda(H,p_i,q_j)$ using
				Lemma~\ref{lem:sp-parallel}\;
			}
			\Else{
				For $i,j\in\{0,1,2,3,4,5\}$, determine $\lambda(H,p_i,q_j)$ using
				Lemma~\ref{lem:sp-series}\;
			}
		}
		$\gamma_{{ds}}(G)=\min\{\lambda(G,x_i,y_j): i,j\in\{0,1,2\}\}$\;
		\Return $\gamma_{{ds}}(G)$\;
	\end{algorithm}
	
	The correctness of Algorithm~\ref{alg:ds-sp} follows from
	Lemmas~\ref{lem:sp-root}, \ref{lem:sp-leaf}, \ref{lem:sp-series}, and
	\ref{lem:sp-parallel}. We now analyze its running time. Assume that $G$ has $n$ vertices and $m$ edges. The parse tree $T$ of $G$ can be constructed in $O(n+m)$ time~\cite{Kikuno1983}. Since $T$ has exactly $m$ leaves and is a binary tree, $T$ has $O(m)$ vertices, and hence the Reverse-BFS ordering $\sigma$ can be computed in $O(m)$ time. At each vertex $v_s$ of $T$, the algorithm determines the values $\lambda(H,p_i,q_j)$, for $i,j\in\{0,1,2,3,4,5\}$,  by Lemmas~\ref{lem:sp-leaf}, \ref{lem:sp-series}, and \ref{lem:sp-parallel}. Each of which is the minimum of a constant number of terms. Thus, each iteration of the ``for'' loop takes $O(1)$ time, and all iterations together take $O(m)$ time. Since a series-parallel graph is
	planar, $m\le 3n-6$, and so Algorithm~\ref{alg:ds-sp} computes
	$\gamma_{{ds}}(G)$ in $O(n)$ time. Therefore, we have the following theorem.
	
	\begin{theorem}\label{thm:sp}
		The dual-server domination number of a given series-parallel graph can be computed in linear time.
	\end{theorem}

	\section*{Declarations}
	
	\noindent{\bf Conflict of interest} The authors do not have financial or non-financial interests that are directly or indirectly related to the work submitted for publication.
	
	\noindent{\bf Data availability}
	No data was used for the research described in this paper.
	
	\section*{Declaration of generative AI and AI-assisted  technologies in the manuscript preparation process}
	
	During the preparation of this manuscript, the authors 
	used OpenAI to improve the language, readability, 
	and presentation of the manuscript. The authors carefully reviewed,  verified, and edited the generated content and take full responsibility for the accuracy and integrity of the published article.

\end{document}